\documentclass[12pt]{iopart}

\expandafter\let\csname equation*\endcsname\relax
\expandafter\let\csname endequation*\endcsname\relax

\usepackage{amsmath}
\usepackage{cite}
\usepackage{xcolor}
\usepackage{graphicx}
\usepackage{braket}

\usepackage{relsize}
\usepackage{blkarray}
\usepackage{url}
\usepackage{ulem}
\usepackage{cancel}
\usepackage{amssymb}

\newcommand{\SI}{Supplemental Material}

\newcommand{\ck}[1]{{#1}}

\begin{document}

\title[Transfer of vibrations via a Rydberg electron]{Transferring vibrational states of trapped atoms via a Rydberg electron}

\author{Abhijit Pendse$^{1}$, Sebastian W\"uster$^2$, Matthew T. Eiles$^{1}$, Alexander Eisfeld$^{1,3,4}$}
\address{$^1$ Max-Planck-Institut für Physik komplexer Systeme, Nöthnitzer Str.\ 38, 01187 Dresden, Deutschland}
\address{$^2$ Department of Physics, Indian Institute of Science Education and Research, Bhopal, Madhya Pradesh 462 066, India}
\address{$^3$ Universität Potsdam, Institut für Physik und Astronomie,
Karl-Liebknecht-Str.\ 24-25, 14476 Potsdam, Deutschland}
\address{$^4$ Institute of Theoretical Physics, TUD Dresden University of Technology, 01062, Dresden, Germany}
\ead{eisfeld@physik.tu-dresden.de}
\vspace{10pt}

\begin{abstract}
We show  theoretically
    that it is possible to coherently transfer vibrational excitation between trapped neutral atoms over a micrometer apart. 
    To this end we consider three atoms, where two are in the electronic ground state and one is excited to a Rydberg state whose electronic orbital overlaps with the positional wave functions of the two ground-state atoms.
    The resulting scattering of the Rydberg electron with the ground-state atoms provides the interaction required to transfer vibrational excitation from one trapped atom to the other.
    By numerically investigating the dependence of the transfer dynamics on the distance between traps and their relative frequencies we find that there is a ``sweet spot" where the transfer of a vibrational excitation is nearly perfect and fast compared to the Rydberg lifetime. 
We investigate the robustness of this scenario with respect to changes of the parameters.
In addition, we derive a intuitive effective Hamiltonian which explains the observed dynamics. 

\end{abstract}

%
%
%
%
%

\section{Introduction}

Recent years have seen tremendous progress in the trapping of neutral atoms \cite{tighttrap_norcia,tighttrap_endres,tighttrap_kaufman,trappedatomcooling_thompson,trappedatomcooling_holzl,atomtrap_kaufman,atomtrap_nogrette}.
It has become possible to trap hundreds of ground-state atoms in nearly arbitrary geometries.  
While the distances between atoms are typically on the order of a few micrometers, the extent of the positional wave function of a trapped atom, which is related to the width of the trap, can be  
smaller than $100\ \mathrm{nm}$ \cite{tighttrap_schlosser,tighttrap_fung}.
Upon excitation to Rydberg states, these atoms can be made to interact through the resulting long-range dipole-dipole interaction even at the length scale set by the trap separation.

     The lifetimes of Rydberg atoms, which can range from micro- to milliseconds with increasing  principal quantum number $\nu$ \cite{BBR_beterov,rydbergtrap_barredo}, are typically long compared to the time scales characteristic of the interactions. 
Therefore, it is not surprising that Rydberg atoms in atomic tweezer arrays have emerged as a versatile and well-controlled platform with potential applications in quantum computation \cite{rydberg_quantcomp_mohan,rydberg_quantcomp_burgers,rydberg_quantcomp_cohen,rydberg_quantcomp_cong,rydberg_quantcomp_saffman_jphysb,rydberg_quantcomp_saffman_rmp,bornet2023scalable,
norcia2024iterative,
rydbergtrap_holz,
unnikrishnan2024coherent,
srakaew2023subwavelength,
chen2023continuous,
scholl2021quantum,
browaeys2020many} and quantum simulation \cite{bernien2017probing,signoles2021glassy,semeghini2021probing,de2019observation,scholl2021quantum,WAtEi11_073044_,ScEiGe15_123005_,Mukherjee_2020,ryd_quant_sim_morgado,ryd_quant_sim_nguyen,ryd_quant_sim_schauss,ryd_quant_sim_scholl,ryd_quant_sim_browaeys}.
Most of the experimental realizations to date have been performed with alkali atoms, which typically become anti-trapped when excited to Rydberg states, although there are exceptions \cite{bhowmik2024double,PhysRevA.98.033411}. However, it has become possible to trap Rydberg atoms in optical potentials by utilizing multivalent (alkaline-earth or lanthanide) atoms or magic wavelengths \cite{rydbergtrap_wilson,rydbergtrap_barredo,rydbergtrap_cortinas}.
Rydberg atoms are also very well suited to investigate long-range resonant excitation transfer via dipole-dipole interactions, which occurs, for example, in natural or artificial light harvesting systems or between  vibrational excitations of subgroups of large molecules \cite{jansen2009waiting,baiz2020vibrational}. 

These applications rely on the long-range Rydberg-Rydberg interaction, and hence, ground-state atoms play no role. 
However, the compatibility between the length scale set by the trap separation and that of the Rydberg orbit enables a second, wholly different, type of long-range interaction between Rydberg and ground-state atoms mediated by the \textit{short-range} electron-atom interaction.
For example, a $\nu=100$ Rydberg atom has an orbital radius of approximately one micrometer, permitting the electronic orbital to overlap several ground-state atoms. 
The interaction between one of these atoms and the Rydberg atom is directly proportional to the electronic density at its position, giving rise to a long-ranged, yet oscillatory, interaction  \cite{fermiSopra1934,greeneCreation2000a,khuskivadzeAdiabatic2002}. 
This leads to a range of remarkable phenomena, ranging from long-range molecule formation in ultracold gases \cite{bendkowskyRydberg2010,rydmol_fey,rydmol_niederprum,peperHeteronuclear2021,boothProduction2015a,eilesTrilobites2019} to, via confinement of atoms in optical potentials \cite{guttridge2023observation,manthey2015}, manipulation of the Rydberg electron in structured environments \cite{eilesAnderson,hunter2020rydberg,PhysRevB.109.075422}.

\begin{figure}
    \centering
   \includegraphics[width=\columnwidth]{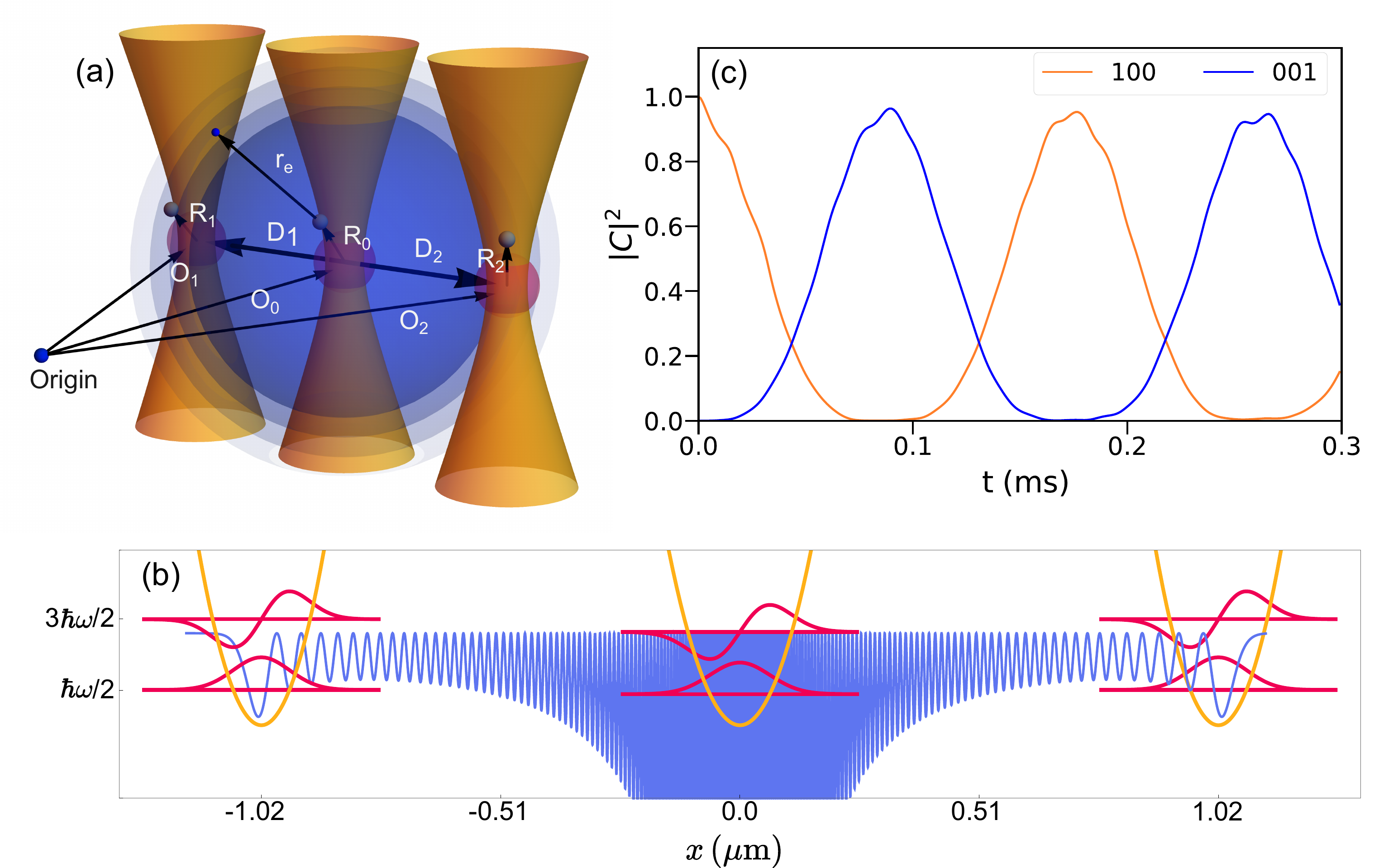}
    \caption{ (a) Sketch of the setup and coordinate system (not to scale). 
    The center of mass coordinates $\vec{R}_i$ of the atoms  are given with respect to the center $O_i$ of the respective trap. 
    The position vector of the Rydberg electron
    $\vec{r}_\mathrm{e}$ is given w.r.t.\ the ion core. The distance vectors between the outer traps and the middle trap  are $\vec{D}_i=\vec{O}_i-\vec{O}_0$.
    (b)  For tight trapping along the y- and z-axes, the system is  effectively one-dimensional. 
    Shown are the ground and first excited vibrational states of the atoms in the traps (red curves) together with the probability density of the Rydberg electron with respect to $\vec{O}_0$ (blue curve). 
    This density is multiplied by $-1$ to illustrate how the interaction between atoms, proportional to this density, becomes attractive when $a_e<0$, leading to molecular formation among other phenomena. The oscillator amplitudes are scaled to appear on the same scale as the Rydberg density, and the $y$ axis is therefore not to scale. The widths of the traps, spatial extent of the harmonic oscillator states, and size of the Rydberg orbit are all to scale. 
    (c) Example of transfer dynamics of vibration from trap 1 to trap 2. 
    The \ck{blue} curve shows the population of the state $\ket{0,0,1}$, where atom 2 is in the first vibrational state of the trap and the other two are in the ground state. 
    The \ck{orange} curve is for the state $\ket{1,0,0}$. 
    The calculations in panel (b) and (c) are done for principal quantum number $\nu=100$, $\omega_1=\omega_2=\omega=2\pi\times 32$\,kHz, trap frequency ratio $\omega_0/\omega = 0.88$, and distances between the traps $D_1=D_2=1.02\mu$m.
}
    \label{fig:setup}
\end{figure}

In this paper, we show that this short-range electron-atom interaction can transfer vibrational excitation from one trap to another.
 The simplest setup to realize this process consists of three traps and is shown in  figure~\ref{fig:setup}\ck{(a)}.
    The two outer traps contain ground state atoms, while the middle one holds a $\nu s$ Rydberg atom,  i.e.\ one with principal quantum number $\nu$ and angular momentum $l=0$.
    The trap separation is such that the positional wave functions of the ground-state atoms can overlap with that of the Rydberg electron of the central atom (figure~\ref{fig:setup}\ck{(b)}).
 An example of vibrational transfer is shown in figure~\ref{fig:setup}\ck{(c)}, which shows the time evolution of the populations of the states $\ket{1,0,0}$ and $\ket{0,0,1}$, where, respectively, the atom of trap 1 (2) is in the first vibrational state and the other two atoms 0, 2 (0, 1) are in the ground state.
 One sees nearly perfect oscillations between the two states on a time-scale of $100\,\mu\mathrm{s}$. 
 All calculations in this paper use parameters corresponding to strontium atoms and a Rydberg $100s$ state, for which we estimate the lifetime to be on the order of a few $100\,\mu\mathrm{s}$ \cite{madjarov2020high,saffmanlifetime,Ybtrapping}. 
 The choice of strontium is motivated by the fact that several groups have setups with trapped strontium atoms in Rydberg states  \cite{rydbergtrap_madjarov,rydbergtrap_holz}, but the general features that we observe will be present in other species as well. 
 The focus of the present work is to understand the underlying process behind this energy transfer and to investigate the dependence of the dynamics on the parameters of the setup, such as the trap frequencies, distances, and Rydberg state.

The paper is organized as follows:
In section \ref{sec:theory} we provide an overview of the theoretical modeling. 
In section \ref{sec:dynamics} we first provide numerical examples of the vibrational transfer dynamics for selected parameters.  We then discuss a simple reduced description which allows us to intuitively understand the numerical observations.
From this reduced description we obtain the parameters for which we expect to see fast and efficient transfer.
This parameter regime is then investigated in detail fully numerically.
Finally, in section \ref{sec:conclusion} we conclude with a discussion of our findings and an outlook.
Details of our derivations and a comparison of the effective Hamiltonian with the full calculations is provided in the \SI{}.

\section{Theoretical description}\label{sec:theory}
 We consider three harmonic traps located at positions $\vec{O}_i$.
    Traps $i=1$ and $i=2$ contain the ground state atoms and trap $i=0$, in the center, the Rydberg atom.
 With $\vec{R}_{i}\equiv(X_{i},Y_{i},Z_{i})$ we denote the center-of-mass coordinates of the $i$th atom with respect to its respective trap centre $\vec{O}_i$.
 With $\vec{r}_\mathrm{e}$ we denote the position of the Rydberg electron with respect to the center-of-mass position $\vec{R}_0$ of its respective core.
     \ck{These coordinates are displayed in  figure \ref{fig:setup}(a).} 
The total Hamiltonian for the center-of-mass degrees of freedom of the three atoms is then given by 
\begin{equation}
H_\mathrm{tot}=\sum_{i=0}^{2}H^\mathrm{(osc)}_{i}+ \sum_{i=1}^{2} H^{\mathrm{(int)}}_{i}.
\label{eq:hamil_split_1}
\end{equation}
Here $H^\mathrm{(osc)}_{i}$ is the Hamiltonian for the trapping of atom $i$ and $H^{\mathrm{(int)}}_{i}$ is the interaction between between states of trap 0 and traps $i=1,2$.
The trapping potential that each atom experiences is given by
\begin{equation}
H^\mathrm{(osc)}_{i}=\frac{\vec{P}^{2}_{i}}{2m_i}+\frac{1}{2}m_i\Big(\omega_{i,X}^{2}X_{i}^{2}+\omega_{i,Y}^{2}Y_{i}^{2}+ \omega_{i,Z}^2 Z_{i}^{2}\Big).
\label{eq:hamil_split_osc}
\end{equation}
where $\vec{P}_i$ is the momentum of atom $i$ and $\omega_{i,X}$, $\omega_{i,Y}$ and $\omega_{i,Z}$ are the trapping frequencies for atom $i$ along the $X$, $Y$ and $Z$ direction, respectively.   
The eigenstates of each oscillator are given by
$
\Psi^{(i)}_{n}(\vec{R}_i)=\Psi^{(i)}_{n}(X_{i},Y_{i},Z_{i})
=\psi^{(i)}_{n}(X_{i})\psi^{(i)}_{0}(Y_{i})\psi^{(i)}_{0}(Z_{i})$,
where for example $\psi^{(i)}_{n}(X_i)$ denotes the  $n$th eigenfunction ($n=0,1,2,\dots$) of a harmonic oscillator with potential $\frac{1}{2}m_i \omega_{i,X}^2 (X_i-O_{i,X})^2 $. 
The corresponding eigenenergies are $E^i_n= \epsilon_i+ n\hbar\omega_{i,X}$
 where $\epsilon_i$ denotes the energy of the ground state.   
 We use as a basis the eigenstates of the non-interacting atoms $\ket{n_1,n_0,n_2}$, which are simply product states of the individual oscillator eigenstates, i.e.,$
     \label{eq:basis_states}
    \braket{\vec{R}_1,\vec{R}_0,\vec{R}_2 |n_1,n_0,n_2} 
    =\Psi^{(1)}_{n_1}(\vec{R}_1)\,\Psi^{(0)}_{n_0}(\vec{R}_0)\,\Psi^{(2)}_{n_2}(\vec{R}_2)$.

The interaction between atom 0 and atom $i$ ($i=1,2)$ is caused by the scattering of the Rydberg electron at position $\vec{O}_0+\vec{R}_0+\vec{r}_\mathrm{e}$ with atom $i$ at position $\vec{O}_i+\vec{R}_i$. 
The electron-atom interaction is described by the Fermi pseudopotential, a zero-range potential whose coupling strength, $U_{e}=2\pi\hbar^{2}a_\mathrm{e}/m_\mathrm{e}$, is proportional to the electron-atom $S$-wave scattering length $a_{e}$  \cite{fermiSopra1934,greeneCreation2000a,fermipseudo_shaffer}.  For Sr, $a_{e}=-13\,a_{0}$, where $a_{0}$ is the Bohr radius \cite{Giannakeas2020a}.  The mass of the electron is $m_\mathrm{e}$. 
Within the Born-Oppenheimer approximation and due to the isotropy of the Rydberg $s$-state, the resulting interaction between atoms in first-order perturbation theory is
\begin{equation}
H^\mathrm{(int)}_{i}=U_\mathrm{e} \; \left|\varphi_{\nu}^{(\mathrm{s})}(\vec{D}_{i}+\vec{R}_{i}-\vec{R}_0)\right|^2.
\label{eq:hamil_split_rydint}
\end{equation}
Here $\vec{D}_i=\vec{O}_i-\vec{O}_0$ is the vector between the position of trap $i$ and trap $0$, 
and $\varphi_{\nu}^{(\mathrm{s})}(\vec r_e)$ is the Rydberg wavefunction.

In the basis of the oscillator product states $\ket{n_1,n_0,n_2}$ the Hamiltonian  (\ref{eq:hamil_split_1}) is written as
\begin{equation}
\begin{split}
    H_\mathrm{tot}=&\sum_{n_1,n_0,n_2} 
    \Big(
    E_0+\sum_{j=0}^3 n_j\, \hbar \omega_j\Big)
    \ket{n_1,n_0,n_2}\bra{n_1,n_0,n_2}
    \\
    &+
    \sum_{{n_1,n_0,n_2}
    \atop{n'_1,n'_0,n'_2}}
    \Big(
 \mathlarger{V}_{(1)}\big[^{n_{0},n_{1}}_{n'_{0},n'_{1}}\big] \, \delta_{n_2,n'_2}
    +
    \mathlarger{V}_{(2)}\big[^{n_{0},n_{2}}_{n'_{0},n'_{2}}\big]\, \delta_{n_1,n'_1}
    \Big) \ket{n_1,n_0,n_2}\bra{n'_1,n'_0,n'_2},
\end{split}
\label{eq:H_0_V_expansion}
\end{equation}
where $E_0$ is the ground state energy of the non-interacting atoms.
The interaction matrix-elements between atom 0 and atom $i$ are 
\begin{equation}
\begin{split}
\mathlarger{V}_{(i)}\big[^{n_{0},n_{i}}_{n'_{0},n'_{i}}\big]
=&
U_\mathrm{e}\int \! \rmd^{3}\vec{R}_{0}\int \!\!\!\rmd^{3}\vec{R}_{i} 
\;
\Psi_{n_{0}}(\vec{R}_0) \Psi_{n_{i}}(\vec{R}_i)\, \Big|\varphi_{\nu}^{(\mathrm{s})}\Big(\vec{D}_{i}+\vec{R}_{i}-\!\vec{R}_{0}\Big)\Big|^{2}
\,
\Psi_{n'_{0}}(\vec{R}_{0}) \Psi_{n'_{i}}(\vec{R}_{i}),
\label{eq:V_i[]_def}
\end{split}
\end{equation}
The matrix-elements are invariant under the exchange $n_0\leftrightarrow n_i$ and  $n'_0\leftrightarrow n'_i$. 
A list of further useful symmetries is given in \ck{Eqs~(S8)} in the \SI{}.

To keep the following discussion most transparent, we consider a co-linear arrangement\footnote{As long as spherically symmetric Rydberg states (i.e.\ s-states) are considered, as  done here,  our treatment is also valid for non-planar arrangements.
} where trap 0 is placed right in the middle between traps 1 and 2.
Furthermore, to simplify numerical calculations, we consider very tight trapping along the $Y$- and $Z$-axes with $\omega_{i,Y},\, \omega_{i,Z}\gg \omega_{i,X}$.
For instance, for the results shown below we have taken $\omega_{i,Y}=\omega_{i,Z}=5\omega_{i,X}$. 
This means that the energetically lowest vibrational states are determined by the trapping frequency $\omega_{i,X}$ along the $X$ axis.
 In this article we focus on the ground and first excited state of each trap.
As we will see below, the strength of the interaction between atom 0 and atom $i$ is roughly one order of magnitude smaller than  $\omega_{i,X}$, so that basis states with energies $\gtrsim 5\,\omega_{i,X}$ can be safely neglected.   
From now on we will denote the trap frequency of atom $i$ along the $x$-axis simply by $\omega_i\equiv \omega_{i,X}$.
These assumptions permit a considerable simplification of the matrix elements in equation~(\ref{eq:V_i[]_def}). \ck{As shown in \SI{} (section S I)}, they reduce to
\begin{equation}
\mathlarger{V}_{(i)}\big[^{n_{0},n_{i}}_{n'_{0},n'_{i}}\big]
\approx U_{e} \iint\rmd x_{0} \;\rmd x_{i}  \;
 \psi_{n_{0}}(x_{0})\;\psi_{n_{i}}(x_{i})\;
\Big|\phi_{\nu}^{(\mathrm{s})}\big(|D_i\!+\!x_{i}\!-\!x_{0}| \big)\Big|^{2}\;
 \psi_{n'_{0}}(x_{0})\;\psi_{n'_{i}}(x_{i}),
\label{eq:T_definition_2}
\end{equation}
where now the integration is only over the $x$ coordinate of each atom position and $\phi_\nu^{(s)}(r_e)$ is the radial electronic wave function.

\section{Dynamics of vibrational excitation transfer}\label{sec:dynamics}

To determine the time evolution, we solve the time dependent Schr\"odinger equation $\rmi\hbar\frac{\partial}{\partial t}\Psi(\vec{R}_{1},\vec{R}_{0},\vec{R}_{2};t)=H_\mathrm{tot}\Psi(\vec{R}_{1},\vec{R}_{0},\vec{R}_{2};t)$. 
We expand the wave function in the basis $\ket{n_{1},n_{0},n_{2}}$  defined above, writing 
$\ket{\Psi(t)}=\sum\limits_{n_{1},n_{0},n_{2}} c_{n_{1},n_{0},n_{2}}(t) \ket{n_{1},n_{0},n_{2}}$.
In the following plots we will focus on the populations $|c_{n_{1},n_{0},n_{2}}(t)|^2$.
As the initial state, we choose the state where atom 1 is initially prepared in the first vibrationally excited state and the other two atoms are in the ground state of their respective traps, i.e., we start in the state $\ket{i}=\ket{n_{1},n_{0},n_{2}}=\ket{1,0,0}$.
We choose a symmetric setup where atom 1 and atom 2 have the same trap frequencies and the corresponding traps have the same distance $D=|D_1|=|D_2|$ from the middle trap 0, where the atom is in the Rydberg state.
Then the matrix elements $\mathlarger{V}_{(1)}\big[^{n_{0},n_{i}}_{n'_{0},n'_{i}}\big]$ and 
$\mathlarger{V}_{(2)}\big[^{n_{0},n_{i}}_{n'_{0},n'_{i}}\big]$ are identical up to a phase (\ck{see equation~(S9) of the  \SI{}}). 
Therefore, in the following we simplify the notation by defining
\begin{equation}
\label{eq:T_definition}
    \mathlarger{T}^{n_{0},n_{1}}_{n'_{0},n'_{i}}\equiv\mathlarger{V}_{(1)}\big[^{n_{0},n_{1}}_{n'_{0},n'_{i}}\big].
\end{equation}
We choose the frequency of the traps to be of the order of a few hundred $\mathrm{kHz}$.
In particular, for all calculations shown, $\omega\equiv \omega_{1,X}=\omega_{2,X} = \ck{2\pi\times32\,\mathrm{kHz}}$, and we vary the trapping frequency of the middle trap.

For the main part of this section we focus on the principal quantum number $\nu=100$, for which the lifetime of the Rydberg state is around a few hundred microseconds and the extent of the Rydberg wave function is roughly one micrometer.
In figure~\ref{fig:setup}\ck(b) we show the Rydberg electron's probability density together with the vibrational wave functions of the atoms in the trap.
From this, one sees that, for the setup considered, the calculation of the matrix elements involves a radial integral predominantly over the outermost lobe of the Rydberg wave function.
The double integral appearing in equation~(\ref{eq:T_definition_2}) is evaluated numerically for the respective parameters.
\ck{We will discuss the dependence on $\nu$ in the conclusions.}
 The time evolution is  obtained  by numerically solving the time dependent Schr\"odinger equation with the Hamiltonian~(\ref{eq:H_0_V_expansion})  in a sufficiently large basis.

\subsection{{Examples of different transfer dynamics}}
\begin{figure}[t]
    \centering
    \includegraphics[width=\columnwidth]{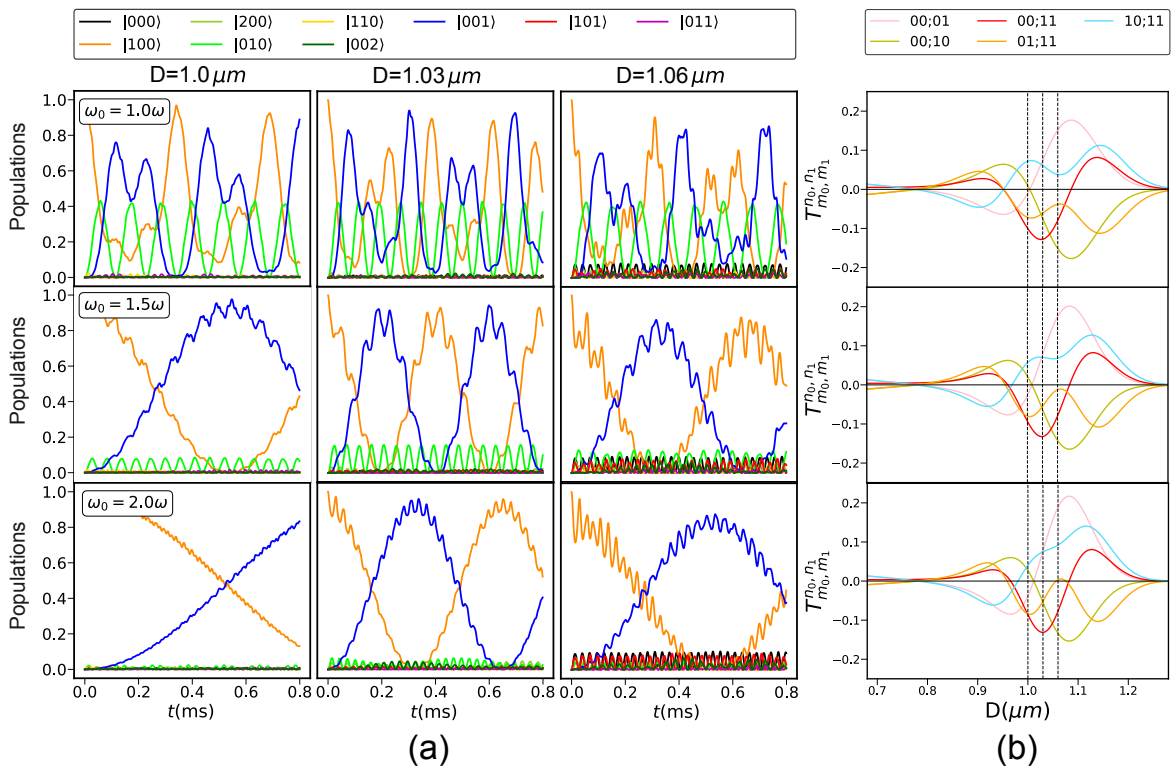}
    \caption{ (a) Population dynamics of the most relevant states and coupling matrix elements.
    In each row a different value of $\frac{\omega_{0}}{\omega}$ is shown; in all panels, $\omega= 2\pi\times 32\,\mathrm{kHz}$.
    In each column a different value of $D$ within the outer lobe of the Rydberg atom is chosen. 
   (b) The separation dependence of the coefficients $\mathlarger{T}^{n_{0},n_{i}}_{n'_{0},n'_{i}}$, defined in equation~(\ref{eq:T_definition}), is shown for the same values of $\omega_0/\omega$ used in (a). 
    The labels in the legend  read $n_{0},n_{i};n'_{0},n'_{i}$. 
    The dashed black vertical lines indicate the values of $D$ used in the different columns of panel (a). 
   }
    \label{fig:examples}
\end{figure}

In  figure~\ref{fig:examples}(\ck{a}) we show the population dynamics for a few selected values of the distance $D$ and the frequency of the middle trap $\omega_0$. 
Each row of \ck{figure~\ref{fig:examples}} corresponds to a different ratio of trap frequencies, and
the different columns of figure~\ref{fig:examples}(\ck{a}) correspond to different $D$ values chosen to be close to the size of the Rydberg orbit $\sim 2\nu^2a_0$.
For $D$ much larger than this size, the Rydberg electron does not overlap the vibrational wave functions and the interaction vanishes. 
On the other hand, for $D$ much smaller, the excited vibrational states of the traps can spatially overlap (c.f. figure~\ref{fig:setup}\ck(b)), depending on the trapping frequency.
In the top row, all three trap frequencies are equal.
For all three distances one sees that the initial vibrational excitation of atom \ck{one} is transferred to atom \ck{two}, but the excitation transfer is never complete: the \ck{blue} curves do not approach a height of one. 
During this process, there is also a strong excitation of vibrational motion of the middle atom \ck{(bright green curve)}.
The time-scale for the transfer from atom 1 to atom 2 (and back) is on the order of a few hundred microseconds.

In the middle row, the Rydberg atom's trap has a higher frequency than the outer traps: $\omega_0=1.5\, \omega$. 
For $D = 1\mu$m separation, there is now a clear oscillation, albeit with small modulations on top, showing vibrational excitation of atom 1 and atom 2 where essentially 100\% of the excitation is transferred. 
There is almost no vibrational excitation of the middle atom. 
Upon increasing the distance $D$, middle and right plot, the time scale of the oscillation first decreases and then increases again. 
Similarly, the middle trap gets excited with a slight decrease of amplitude for the larger value of $D$. 
In the bottom row, $\omega_0$ is increased even further to $\omega_0=2\, \omega$.
The dynamics of the main oscillation becomes even slower than before, and
excitations of the middle trap are negligible.

The results shown here illustrate that, in general, the dynamics has a complicated and strong dependence on the distance $D$ and frequency difference $\omega_0-\omega$.
 In general, for $\omega\sim\omega_0$, very complex dynamics ensue and full excitation transfer does not occur. As $\omega_0 -\omega$ grows, the dynamics tend to simplify. 
For example, we found that for $\omega_0\gtrsim 2\, \omega$ clear Rabi-like oscillations (cf.\ last row of figure~\ref{fig:examples}) occur. These demonstrate  nearly 100\% transfer of vibrational excitation from trap 1 to trap 2.
Since the time scale shown in figure~\ref{fig:examples} is comparable to the lifetime of the Rydberg state, we aim for a parameter regime where the excitation transfer is as fast as possible, while still yielding as close as possible to 100\% transfer. 
One such optimal parameter regime yields the results shown in figure~\ref{fig:setup}\ck{(c)}. 
To understand the parameter dependence of this optimal scenario, we introduce a simple analytical model in the next subsection \ck{\ref{sec:analytics}} which gives clear insight. 

We conclude this subsection by taking a closer look at the dependence of the matrix elements $\mathlarger{T}^{n_{0},n_{i}}_{n'_{0},n'_{i}}$ on $D$ and $\omega_0$.
In \ck{figure~\ref{fig:examples}(b)},  we show the values of these matrix elements as a function of $D$ for the same values of $\omega_0$ as in the corresponding rows of \ck{figure~\ref{fig:examples}(a)}.  
Shown are the most relevant elements for our initial condition where we only have one vibrational excitation.
The first thing to note is that the matrix elements are mostly insensitive to the frequency $\omega_0$.
Furthermore, we see that the values of the matrix elements are small compared to the frequencies of the traps, which will be advantageous in our analytical consideration presented in the \ck{next subsection}.
We note that matrix elements between the states with one excitation and those with larger excitation numbers have a complicated distance dependence and typically decrease only slowly for increasing  excitation. 
From the analytical considerations in the following subsection it becomes clear that these states nevertheless become less relevant because of the increasing energy separation.

We see that the matrix elements change strongly over the range of $D$ shown, which is around $500\,\mathrm{nm}$.
As discussed earlier, all matrix-elements vanish when $D\gtrsim 1.3\,\mu\mathrm{m}$, because the distance is so large that there is no overlap of the wave function of the Rydberg electron with the ground state atom.
Likewise, 
we do not show distances smaller then $D<0.8\,\mu\mathrm{m}$, since for such small distances overlap of oscillator wave functions of different traps can occur when the wave functions are in the third excited state or higher.

\subsection{Effective state models\label{sec:analytics}}

 In this subsection we provide some simple analytical considerations that help to understand the dynamics observed in figure \ref{fig:examples}, and guide our search for suitable parameters for optimal state transfer.
 The results of figure \ref{fig:examples} indicate that the most relevant states for the dynamics are those that have the same energy (or nearly the same energy) as the initial state.
 Due to the symmetry of our setup the state $\ket{0,0,1}$ has the same energy as the initial state $\ket{1,0,0}$.
 When, additionally, $\omega_0=\omega$, the state $\ket{0,1,0}$ has the same energy and strongly participates in the dynamics.
 In contrast, when the frequency of the middle trap is much larger than that of the outer ones (e.g.\ $\omega_0=2\,\omega$ as in the bottom row of figure \ref{fig:examples}), then the $\ket{1,0,0}$ and $\ket{0,0,1}$ are the only relevant states.
  The cases in between are more complicated and depend on the detailed parameters.

We now construct effective Hamiltonians in the respective state spaces.
To this end, we employ a second-order treatment in the interaction.
Such a treatment yields good results when the interaction is small compared to the energy difference between those states included in the relevant subspace and those outside.
\ck{figure~\ref{fig:examples}(b)} confirms that the relevant matrix elements $\big|\mathlarger{T}^{n_{0},n_{1}}_{n'_{0},n'_{i}}\big|$ are $\lesssim 0.2\,\hbar \omega$.

\subsubsection{$\omega_0\gg \omega$ scenario:}
In this case, the relevant state space is spanned by the states  $\ket{1,0,0}$ and $\ket{0,0,1}$.
As shown in \ck{S.II of the SI{}} one finds for the effective Hamiltonian (neglecting higher order terms in $\omega/\omega_0$) 
\begin{equation}
 H_{\mathrm{eff}}^{\mathrm{(2-state)}}\approx
\begin{pmatrix}
0 & \kappa\\
\kappa & 0
\end{pmatrix}
\qquad
\mathrm{with}
\qquad
\frac{\kappa}{\hbar \omega_0}
\approx
2\sum\limits_{j=1}\frac{(-1)^{j}}{j}\,
\Bigg(\frac{{\mathlarger{T}_{j,1}^{0,0}} }{\hbar \omega_0}\Bigg)^2.
\label{eq:2state_effective_hamiltonian}
\end{equation}
This implies that, when the system is initially prepared in the state $\ket{1,0,0}$, the populations of this state and $\ket{0,0,1}$ undergo Rabi oscillations with frequency $\kappa/\hbar$.
Such population dynamics are clearly shown in the bottom row of figure~\ref{fig:examples}.
We have found that for the distances $D$ shown in the figure the dominant contribution to $\kappa$ stems from  $T^{0,0}_{1,1}$ .  
As $D$ increases from $1.00\,\mu$m to $1.06\,\mu$m, the magnitude of this matrix element first increases, then decreases, leading to the change in frequency clearly reflected in figure~\ref{fig:examples}(a).

Although  for $\omega_0=2\, \omega$ the detuning is not very large, we nevertheless find qualitative agreement between the dynamics of the full Hamiltonian and that of the effective Hamiltonian, as can be seen in \ck{figures~S1 and S3 of the \SI{}}.
There, one also sees that for $D=1.06\,\mu\mathrm{m}$, the results of the effective Hamiltonian differ substantially from the full results. 
This is expected, since in the respective panel of figure~2\ck{(a)},  many states outside of the reduced subspace are populated.

\subsubsection{$\omega\sim\omega_0$ scenario: \label{sec:omega0=omega}}
We now examine the scenario when all trap frequencies are similar.
The relevant subset of quasi-degenerate states is spanned by the states $\ket{1,0,0}$, $\ket{0,1,0}$, and $\ket{0,0,1}$. 
To first order in the interaction, we have
\begin{equation}
 H_{\mathrm{eff}}^\mathrm{(3-state)}\approx
\begin{pmatrix}
0 &\alpha&0\\
\alpha & \Delta & \alpha \\
0 & \alpha & 0 &
\end{pmatrix},
\qquad
\mathrm{with}
\begin{array}{rl}
&\Delta=\hbar(\omega_{0}-\omega)
-\alpha ,\\
\\
&\alpha= \mathlarger{T}^{0,0}_{1,1}.
\end{array}
\label{eq:3state_effective_hamiltonian_1}
\end{equation}
The derivation is given in \ck{section S~II.2  of the \SI{}}, which also details higher-order terms.
In the \ck{\SI{} (section S III)} we show that there is  good agreement between the full numerical dynamics and that stemming from the effective Hamiltonian.
The agreement is generally better for smaller $\omega_0$ and for distances $D$ around $1.03\,\mu\mathrm{m}$.

For the special case of $\Delta=0$, the dynamics resulting from $H_{\mathrm{eff}}^\mathrm{(3-state)}$ is particularly simple.
The probability to find the system in the state $\ket{0,0,1}$ after initial preparation in the state $\ket{1,0,0}$ is 
$p(t)= \frac{1}{4}\big(\cos(\frac{\sqrt{2}\alpha t}{\hbar})-1\big)^2$.
This implies that the vibrational excitation has been completely transferred from trap 1 to trap 2 at the time $t_0=\frac{\pi \hbar}{\sqrt{2} \alpha}$. 
When $\Delta \ne 0$, the dynamics is more complicated and typically resembles that seen in \ck{the upper row of figure~\ref{fig:examples}\ck{(a)}}.

\subsubsection{Discussion:}
As noted above, both analytical formulas (\ref{eq:2state_effective_hamiltonian}) and (\ref{eq:3state_effective_hamiltonian_1}) indicate that the system dynamics is mainly determined by the interaction element $\mathlarger{T}^{0,0}_{1,1}$.
When the middle oscillator is significantly detuned from the other oscillators, the analytical results from above show that the characteristic timescale of the population transfer is given by the square of this (small) matrix element.
On the other hand, when all traps have three similar frequencies, the characteristic timescale is linear in this quantity.
This is consistent with the observation from figure~\ref{fig:examples} that the transfer dynamics slows down upon increasing the detuning.
Therefore, in what follows we will focus on the case where all three traps have similar frequencies.
While the analytic expressions from above can serve as a guideline, in the end one has to perform full numerical simulations.
As remarked above, for $\Delta=0$ the analytical dynamics gives 100\% transfer of vibrational excitation.
From equation~(\ref{eq:3state_effective_hamiltonian_1}) we see that $\Delta=0$ implies $\hbar\omega_0=\hbar\omega +\mathlarger{T}^{0,0}_{1,1} $, which means that we need a (small) detuning of the middle oscillator to achieve this scenario.
From the red curves of \ck{figure~\ref{fig:examples}(b)} we see that this detuning is negative for distances $D$ \ck{roughly between 0.9 and $1.1\,\mu\mathrm{m}$} and positive outside that interval. 
The largest absolute value of $\mathlarger{T}^{0,0}_{1,1}$  in this interval occurs for separations $D\approx 1.00\, \mu \mathrm{m}$.
Therefore, in the next \ck{subsection} we consider this regime in more detail.

\subsection{Fast and efficient transfer and its robustness against variation of parameters\label{sec:optimal_trans}}
\begin{figure}
    \centering
    \includegraphics[width=\columnwidth]{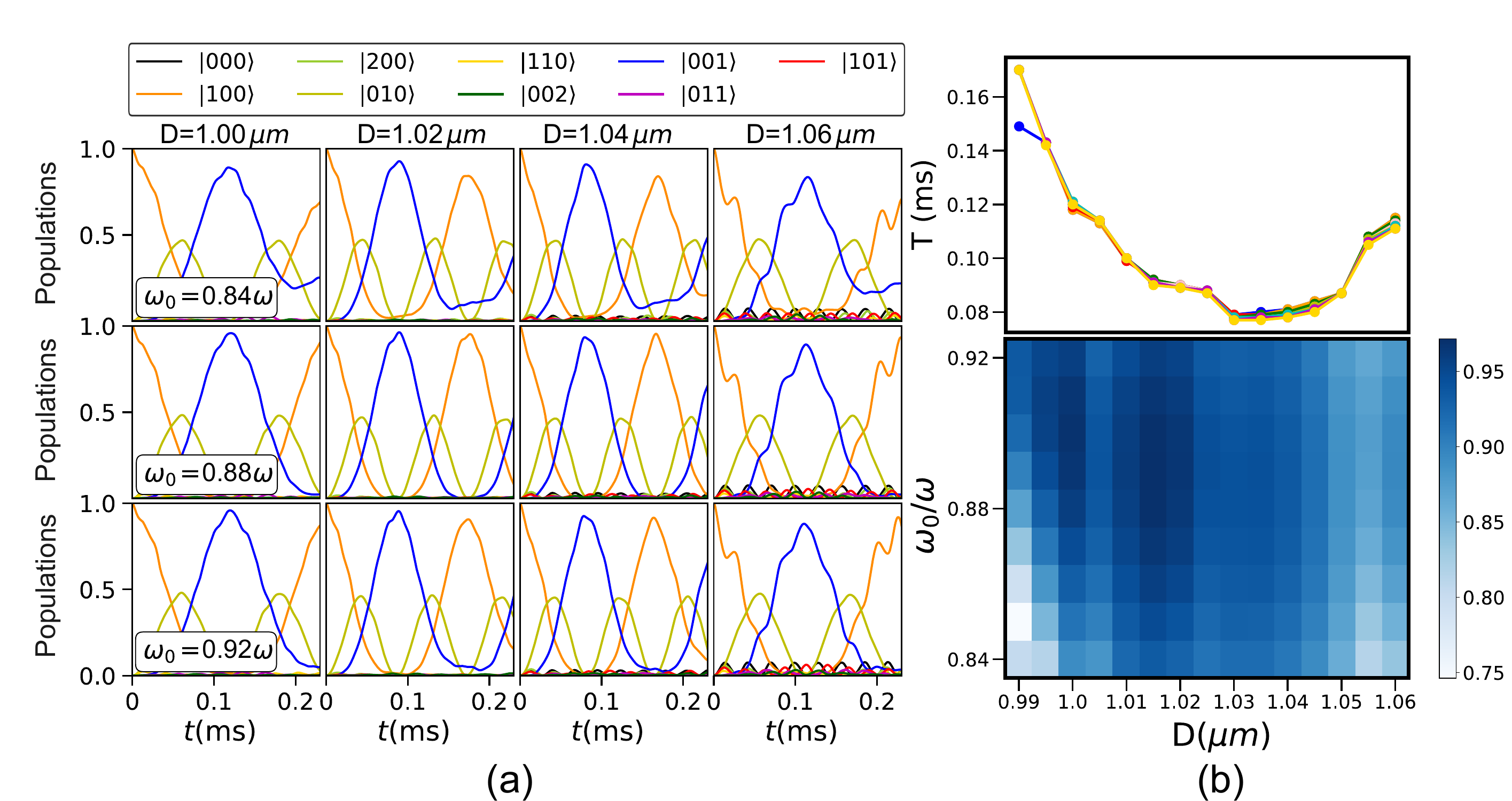}
    \caption{
    Robustness of the fast state transfer of figure~\ref{fig:setup}\ck{(c)} against variations in $\omega_0$ and $D$.
    (a) Population dynamics. Each row is for a fixed $\omega_0$ (provided in the figure) and contains dynamics for 4 different distances $D$ (given in each panel).
    \ck{(b)} \ck{Top}: Time $T$ when the vibrational excitation probability $P_2(t)$ on trap 2 reaches its first maximum as function of $D$. 
    \ck{Bottom}: The population corresponding to this maximum.
    }
\label{fig:Optimal_trimer}
\end{figure}

Figure~\ref{fig:Optimal_trimer}\ck{(a)} shows the time-dependent populations for several values of $D$ and $\omega_0$ chosen such that the resulting dynamics is similar to that shown in figure~\ref{fig:setup}\ck{(c)}.
For all $\omega_0$, the quantity $\Delta$ defined in equation~(\ref{eq:3state_effective_hamiltonian_1}) \ck{(see also equation~(S13) of the \SI)} is very small such that the basic dynamics can be understood by the discussion at the end of section \ref{sec:omega0=omega}.
In all panels with $D\ge 1.0\,\mu\mathrm{m}$, the population of the state $\ket{1,0,0}$ is close to zero while the population of the state $\ket{0,0,1}$ is nearly one at $t\approx 0.1\, \mathrm{ms}$.
For $D=1.00\,\mu\mathrm{m}$, the time at which the population in the right trap reaches its first maximum is considerably larger than $0.1\,\mathrm{ms}$; for even smaller $D$ it grows quickly (e.g., for $D=0.98\,\mu\mathrm{m}$ it is twice as large as for $D=1.02\,\mu\mathrm{m}$).
We also see that for the first row ($\omega_0=0.84\,\omega$) the maximum of the first peak of the population of $\ket{0,0,1}$ is smaller than the corresponding populations of the second row. 
Similarly, in the last row the populations of the first peak is (for the cases with $0.1\, \mu\mathrm{s}$) smaller than the corresponding ones of the middle row.
The optimal scenario - by which we mean that the transfer is fast and the population transfer becomes large - is shown in the panel with $D=1.02\,\mu\mathrm{m}$ and $\omega_0=0.88\,\omega$.
To investigate these two criteria for the optimal scenario further, we have calculated the `time of the peak' $T$ and the corresponding population at the peak maximum for a larger range of values and a finer spacing.
The findings are presented in figure~\ref{fig:Optimal_trimer}\ck{(b)} where the  maximum of the first peak in the population of the state $\ket{0,0,1}$ as a function of $D$ and $\omega_0$ is shown (bottom) together with the corresponding time at which this maximum appears (top). 
Over the range of $\omega_0$ values shown  ($0.84 \leq \omega_0/\omega \leq 0.92 $)  all curves look essentially the same as function of $D$.
In the range $1.01\,\mu\mathrm{m}\lesssim D\lesssim 1.04\, \mu\mathrm{m}$ there is a broad minimum where the first peak appears at around $t\approx0.09\,\mathrm{ms}$.
Outside this range of distances $D$, the time of the first maximum quickly increases.
We note that outside the range shown (starting already at the borders), the assignment of a first maximum becomes slightly arbitrary, because of the fast oscillatory behaviour on top of a slow transfer oscillation. 
When we look at the amplitude of the first peak in \ck{panel d}, then we see that there is a broad region where the amplitude of the first peak is close to one (dark blue region).
There is a `sweet spot' around (\ck{$D\approx 1.02\,\mu\mathrm{m}$}, \ck{$\omega_0\approx 0.88\,\omega$}) where 
the amplitude of the first peak is close to one and the time is close to the minimum.
These values of $\omega_0$ and $D$ have been chosen for the results shown in  figure~\ref{fig:setup}\ck{c}.
From the results of figure~\ref{fig:Optimal_trimer}\ck{(b)} one see that this `sweet spot' lies in the center of a region where the amplitude and the time of the first maximum are nearly unchanged. 
This region has a width of approximately $30\,\mathrm{nm}$ and $0.05\,\omega$.
That means one needs an accuracy in placing the traps of about $15\,\mathrm{nm}$ and an accuracy in the trap frequencies of about $5\%$.

\section{Conclusions}\label{sec:conclusion}

In this article, we have investigated the possibility that the vibrational excitation of trapped neutral atoms can be transferred from one trap to another, even across micrometer-scale distances, using the scattering of a Rydberg electron off of the neutral atoms. 
 We have focused on the situation where the traps are arranged collinearly and where one of the outer atoms is initially prepared in the first vibrationally excited trap state with the other traps prepared in the ground state.
We found a parameter range for which such transfer is possible, exploring in detail the dependence on the distance of equidistant traps and on the frequency detuning of the middle trap. 
There remain additional parameter choices that we did not explore, for example the principle quantum number of the Rydberg state, the frequencies of the outer traps, or the linear symmetric arrangement around the middle trap.
In the following, we briefly comment on these choices in connection with experimental challenges.

\textit{The principle quantum number:}
The spatial extent of the Rydberg state scales as $\nu^2$.
To obtain large interaction matrix elements, the outer lobe of the Rydberg wave function must significantly overlap the vibrational wave functions of the outer traps. 
Thus, the distance between traps determines the minimum value of $\nu$, and in general we must select $D\sim 2\nu^2$ for maximal overlap for a given $D$. 
However, the magnitude of the interaction matrix elements decrease rapidly with $\nu$, setting an upper limit on $\nu$ and therefore $D$. 
We estimate that the interaction matrix elements, provided $D\sim 2\nu^2$, scale approximately as $\nu^{-6}$. 
This scaling is a result of the normalization constant of the Rydberg wave function, $\nu^{-3}$.
 The lifetime of the Rydberg state also depends strongly on the principle quantum number $\nu$.
It scales either as $\nu^2$ or $\nu^3$, depending on the relevance of black-body radiation.
Since the transfer timescale is inversely proportional to the interaction, we see that smaller $\nu$ values provide a better ratio of transfer time to Rydberg lifetime. 
That means that the distance between the traps should be as small as possible.
Our chosen value of the distance $D_\mathrm{min}\approx 1\, \mu\mathrm{m}$ should be realizable with present day setups.
This distance then leads to $\nu \sim 100$. 
We note that Rydberg dressing might provide a way to better optimize these parameters \cite{PhysRevA.65.041803,Balewski_2014,PhysRevLett.116.243001}.

\textit{The trapping frequencies:}
The chosen value of $\omega_x=2\pi \times 32\,\mathrm{kHz}$  is selected such that perpendicular trapping frequencies of around $4\,\omega_x$ would be experimentally achievable.
    We note that the tight trapping in the $y$ and $z$ axes was assumed in order to have one-dimensional oscillators, which both facilitates the numerical investigation and also avoids degeneracies in the individual traps, allowing us to obtain a clear picture of a single vibrational excitation.
In general, we expect that tighter trapping in the $x$ direction is favorable.
We stress that tight trapping in the $y$ and $z$ axes is not essential to observe vibrational excitation transfer, but we expect that the resulting dynamics becomes more complicated otherwise.
Our investigations in \ck{section \ref{sec:optimal_trans} } show that the transfer dynamics do not change appreciably upon small variation of $D$ and $\omega$. 
To remain within the ideal window of optimal parameter requires an accuracy in the trap frequency of about $5\%$ and an accuracy of placing the traps of about $15\,\mathrm{nm}$.
Such accuracy is at the edge of what is presently achievable \cite{trapplacement_visscher,trapplacement_catala}.

\textit{Symmetry of the arrangement:} The assumption of equal frequencies of the outer traps and equal distances between the traps was  primarily made for convenience.
For more asymmetric arrangements, we expect that the efficient transport will be impeded.
The linearity of the setup is also made mostly for convenience, since for a Rydberg $s$-state the interaction depends only on the distance between Rydberg and ground state atom. 

\textit{Choice of atomic species:}
As already mentioned, we have chosen strontium, as it is currently used in several experiments involving trapped Rydberg atoms.  
However, other atoms that can be trapped in Rydberg states can also be used, which may improve the lifetime or strength of the interaction through its dependence on the electron-atom scattering length. 
For example, $a_\mathrm{e} \approx-20\,a_0$ for Cs \cite{eilesTrilobites2019}, which would increase the speed of the dynamics by almost a factor of two. 

In the present work we have restricted the discussion to the populations of the vibrational states.
It would be interesting also to investigate the entanglement properties. 
For example, figure~\ref{fig:examples} shows that, for detunings $\omega_0 \gtrsim 1.5\,\omega$, the system can be in a Bell state $\frac{1}{\sqrt{2}}(\ket{1,0,0}\pm \ket{0,0,1})$.
Since the dynamics is slow for such a large detuning, one would have to perform simulations taking decoherence processes such as the decay of the Rydberg atom into account to adequately study entanglement in this scenario. 
We also restricted our discussion to initial states where only one vibrational excitation is present.
This was done primarily to keep the basis in our numerical calculations small and to provide a simple and clear picture of the transfer dynamics.
It would be interesting to investigate the dynamics when initially one trap is excited to higher vibrational states, for example coherent oscillator states.
Furthermore, it would be interesting to investigate transport through larger networks of trapped atoms, where for example every other atom is excited to a Rydberg state.

\ack
A.E.\ acknowledges support from the Deutsche Forschungsgemeinschaft (DFG) via a Heisenberg fellowship (Grant No EI 872/10-1)

\newpage


\bibliographystyle{journal_v5}

\bibliography{Alex,references_3atom}
%

\end{document}


\title[]{\center Supplementary Material 
\\
--------------------------------------
\\
Transferring vibrational states of trapped atoms via a Rydberg electron}

\author{Abhijit Pendse$^{1}$, Sebastian W\"uster$^2$, Matthew T. Eiles$^{1}$, Alexander Eisfeld$^{1,3,4}$}
\address{$^1$ Max-Planck-Institut für Physik komplexer Systeme, Nöthnitzer Str.\ 38, 01187 Dresden, Deutschland}
\address{$^2$ Department of Physics, Indian Institute of Science Education and Research, Bhopal, Madhya Pradesh 462 066, India}
\address{$^3$ Universität Potsdam, Institut für Physik und Astronomie,
Karl-Liebknecht-Str.\ 24-25, 14476 Potsdam, Deutschland}
\address{$^4$ Institute of Theoretical Physics, TUD Dresden University of Technology, 01062, Dresden, Germany}
\ead{eisfeld@physik.tu-dresden.de}

\date{\today}

\vspace{40 pt}

\noindent
In this \SI{} we provide additional details on the following topics:
\vspace{1cm}

\begin{minipage}{16cm}
\begin{itemize}
\item[S\,I.] Derivation of matrix elements for the case of tight perpendicular trapping.
\item[S\,II.] The elements of the effective Hamiltonians.

\item[S\,III.]
Comparison of the dynamics of the effective and the full Hamiltonian. 
\end{itemize}
\end{minipage}

\newpage

\section{Derivation of the matrix elements for tight trapping }\label{app:EOM_derivation}

We start with 
\ck{Eq.~(5)}.
For tight trapping along the $Y$ and $Z$ axes the relevant states of atom $i$ are given 
as $\Psi^{(i)}_{n}(\vec{R}_i)=\Psi^{(i)}_{n}(X_{i},Y_{i},Z_{i})
=\psi^{(i)}_{n}(X_{i})\psi^{(i)}_{0}(Y_{i})\psi^{(i)}_{0}(Z_{i}) $. 
We now insert these product states into \ck{Eq.~(5)} and use the explicit form of the ground state wavefunction $\psi^{(i)}_{0}(\zeta)= \Big(\frac{m\Omega}{\pi\hbar}\Big)^{\frac{1}{4}}\; e^{-\frac{m\Omega\zeta^{2}}{2\hbar}}$ for harmonic potentials with frequency  $\Omega$. 
We then have\footnote{
Here, the function
$
\phi_{\nu}^{(\mathrm{s})}\Big(|\vec{r}|\Big)= \; \sqrt{\frac{1}{\pi a_{0}^{3}}\Big(\frac{1}{\nu}\Big)^{5}}  \; \frac{e^{-\frac{r}{\nu a_{_{0}}}}}{r} \;  L^{^{1}}_{_{(\nu-1)}} \Big(\frac{2r}{a_{_{0}}\nu}\Big)
$,
where $a_{0}$ is the Bohr radius and $L^{^{1}}_{_{(\nu-1)}}$ is the generalized Laguerre polynomial of degree $(\nu-1)$, is the result of transforming the spherically isotropic $s$-wave radial wave function in spherical coordinates into cartesian coordinates. 
.}
\begin{equation}
\begin{split}
\mathlarger{V}_{(i)}\big[^{n_{0},n_{i}}_{n'_{0},n'_{i}}\big]
=&
U_\mathrm{e}
 \iint \! \rmd x_{0} \rmd x_{i} \,
 f^{n_0,n_i}_{n'_0,n'_i}(x_0,x_i)
  F(x_0,x_i)\; 
\label{eq:V_i[]_appendix_1}
\end{split}
\end{equation}
with 
\begin{equation}
\begin{split}
    f^{n_0,n_i}_{n'_0,n'_i}(x_0,x_i)=&\ 
    \psi_{n_{0}}(x_{0})\psi_{n_{i}}(x_{i})
    \,\psi_{n'_{0}}(x_{0})\psi_{n'_{i}}(x_{i})
\\
\\
F(x_0,x_i)=&\int\! \rmd y_{0}\rmd z_{0}\rmd y_{i}\rmd z_{i}\;K(x_0,y_0,z_0,x_i,y_i,z_i )\cdot\Big(\frac{m\Omega}{\pi\hbar}\Big)^{2}\; e^{-\frac{m\Omega(y_{0}^{2}+z_{0}^{2})}{\hbar}}\; e^{-\frac{m\Omega(y_{i}^{2}+z_{i}^{2})}{\hbar}}
\end{split}
\end{equation}
with $ K(x_0,y_0,z_0,x_i,y_i,z_i )=\Big|\phi_{\nu}^{\mathrm{(s})}\big(\sqrt{(D_i+\!x_{i}\!-\!x_{0})^2+(y_{i}\!-\!y_{0})^2+(z_{i}\!-\!z_{0})^2}
\big)\Big|^{2} $.
Here we have introduced $D_1=-D$ and $D_2=D$ which is valid when both outer traps are equidistant from the central trap.
Next, we transform to new coordinates, $y_\mathrm{c}=y_i+y_0$, $y_\mathrm{r}=y_i-y_0$, $z_\mathrm{c}=z_i+z_0$, $y_\mathrm{c}=z_i+z_0$, to obtain
\oAlex{
\begin{equation}
\begin{split}
F(x_0,x_i)
\mathlarger{V}_{(i)}\big[^{n_{0},n_{i}}_{n'_{0},n'_{i}}\big]
=&
U_\mathrm{e}
 \iint \! \rmd x_{0} \rmd x_{i} \frac{1}{4}\iint\rmd y_\mathrm{r}\rmd z_\mathrm{r}\; \Bigg[ \psi_{n_{0}}(x_{0})\psi_{n_{i}}(x_{i})\,\psi_{n'_{0}}(x_{0})\psi_{n'_{i}}(x_{i}) 
 \\
& \qquad\qquad \cdot\Big|\phi_{\nu}^{\mathrm{(s})}\big(\sqrt{(D_i+\!x_{i}\!-\!x_{0})^2+y_\mathrm{r}^2+z_\mathrm{r}^2}\big)\Big|^{2}\\
&\hspace{5cm}\cdot\Big(\frac{m\Omega}{\pi\hbar}\Big)^{2}\; e^{-\frac{m\Omega(y_\mathrm{r}^{2}+z_\mathrm{r}^{2})}{\hbar}}\; \iint \rmd y_\mathrm{c}\rmd z_\mathrm{c}e^{-\frac{m\Omega(y_\mathrm{c}^{2}+z_\mathrm{c}^{2})}{ \hbar}}\Bigg].
\label{eq:V_i[]_appendix_2}
\end{split}
\end{equation}
}
\begin{equation}
\begin{split}
F(x_0,x_i)
=&
 \frac{1}{2}  \iint\rmd y_\mathrm{r}\rmd z_\mathrm{r}\; \Bigg[ \Big|\phi_{\nu}^{\mathrm{(s})}\big(\sqrt{(D_i+\!x_{i}\!-\!x_{0})^2+y_\mathrm{r}^2+z_\mathrm{r}^2}\big)\Big|^{2}\cdot\Big(\frac{m\Omega}{\pi\hbar}\Big)\; e^{-\frac{m\Omega(y_\mathrm{r}^{2}+z_\mathrm{r}^{2})}{\cabp{2}\hbar}}\; \Bigg],
\label{eq:V_i[]_appendix_3}
\end{split}
\end{equation}
where we have used  $\iint \rmd y_\mathrm{c}\rmd z_\mathrm{c}e^{-\frac{m\Omega(y_\mathrm{c}^{2}+z_\mathrm{c}^{2})}{2\hbar}}= {2\frac{\pi\hbar}{m\Omega}}$. 
Now we transform $y_\mathrm{r}$ and $z_\mathrm{r}$ to polar coordinates 
\begin{equation}
\begin{split}
F(x_0,x_i)
=&
\Bigg[  \frac{1}{2} \int_0^{2\pi}\! \rmd \varphi \int_0^\infty \! \rmd \rho\,\rho\,\Big|\phi_{\nu}^{\mathrm{(s})}\big(\sqrt{(D_i+\!x_{i}\!-\!x_{0})^2+\rho^2}\big)\Big|^{2}\cdot\Big(\frac{m\Omega}{\pi\hbar}\Big)\; e^{-\frac{m\Omega }{\cabp{2}\hbar}\rho^2}\; \Bigg].
\label{eq:V_i[]_appendix_4}
\end{split}
\end{equation}
 Since all the integrands are independent of $\varphi$, one obtains
 \oAlex{
\begin{equation}
\begin{split}
\mathlarger{V}_{(i)}\big[^{n_{0},n_{i}}_{n'_{0},n'_{i}}\big]
=&
U_\mathrm{e}
 \iint \! \rmd x_{0} \rmd x_{i} \Bigg[ \psi_{n_{0}}(x_{0})\psi_{n_{i}}(x_{i})\,\psi_{n'_{0}}(x_{0})\psi_{n'_{i}}(x_{i}) 
 \\
& \qquad \cdot\frac{1}{2} \int_0^\infty \! \rmd \rho\,\rho\,\Big|\phi_{\nu}^{\mathrm{(s})}\big(\sqrt{(D_i+\!x_{i}\!-\!x_{0})^2+\rho^2}\big)\Big|^{2}\cdot\Big(\frac{m\Omega}{\hbar}\Big)\; e^{-\frac{m\Omega }{\hbar}\rho^2}\; \Bigg].
\label{eq:V_i[]_appendix_5}
\end{split}
\end{equation}
 Introducing $l_{_{\Omega}}=\sqrt{\frac{\hbar}{m\Omega}}$, the oscillator length for transverse trapping with frequency $\Omega$, and scaling the integration variable $\rho$ one obtains 
 }
 \begin{equation}
\mathlarger{V}_{(i)}\big[^{n_{0},n_{i}}_{n'_{0},n'_{i}}\big]
=U_{e}\; \int\limits_{-\infty}^{}\!\!\!\!\!\int\limits_{}^{\infty} \rmd x_{0} \;\rmd x_{i}  \;
 \psi_{n_{0}}(x_{0})\;\psi_{n_{i}}(x_{i})\;
 S_{(i)}(x_i-x_0)\;
 \psi_{n'_{0}}(x_{0})\;\psi_{n'_{i}}(x_{i}),
\label{eq:T_definition}
\end{equation}
with
\begin{equation}
S_{(i)}(x_i-x_0)=
\int_{0}^{\infty}\!\! \rmd\rho \;\; \Big|\phi_{\nu}^{(\mathrm{s})}\big(\sqrt{(D_i\!+\!x_{i}\!-\!x_{0})^2+(l_\Omega\rho)^2} \big)\Big|^{2} \rho \, e^{-\frac{\rho^{2}}{2 }}.
\end{equation}
Here, $l_{_{\Omega}}=\sqrt{\frac{\hbar}{m\Omega}}$ is the oscillator length corresponding to the trapping frequencies along the Y and Z axes].
In the region of overlap of the Rydberg electron with the  center of mass wave function of the other atoms, the Rydberg wavefunction can approximately be taken as constant in the directions perpendicular to the $x$-axis (because of tight trapping) and we obtain

$
S_{(i)}(x_i-x_0)\approx
\Big|\phi_{\nu}^{(\mathrm{s})}\big(\sqrt{(D_i\!+\!x_{i}\!-\!x_{0})^2} \big)\Big|^{2} \, \int_{0}^{\infty}\!\! \rmd\rho \;\rho  e^{-\frac{\rho^{2}}{2 }}
$

and therefore
\begin{equation}
S_{(i)}(x_i-x_0)\approx
\Big|\phi_{\nu}^{(\mathrm{s})}\big(\sqrt{(D_i\!+\!x_{i}\!-\!x_{0})^2} \big)\Big|^{2} 
\label{eq:S_simp_2}
\end{equation}
which leads to the form given in \ck{Eq.~(6)}.
\vspace{1.5cm}

\subsection{Symmetries of the interaction matrix elements}
 One easily sees the following  symmetries from Eq.(6) of the main text, 
 \begin{equation}
\mathlarger{V}_{(i)}\big[^{n'_{0},n'_{i}}_{n_{0},n_{i}}\big]=\mathlarger{V}_{(i)}\big[^{n'_{0},n_{i}}_{n_{0},n'_{i}}\big]=\mathlarger{V}_{(i)}\big[^{n_{0},n'_{i}}_{n'_{0},n_{i}}\big]=\mathlarger{V}_{(i)}\big[^{n_{0},n_{i}}_{n'_{0},n'_{i}}\big].
\label{eq:ST_symmetry1_apndx}
\end{equation}

\noindent
Furthermore, for  $|D_1|=|D_2|$ we have a simple relation between matrix elements of $V_{(1)}=V_{(2)}$:
\begin{equation}
    \mathlarger{V}_{(1)}\big[^{n_{0},n}_{n'_{0},n'}\big]=(-1)^{(n'_{0}+n'+n_{0}+n)} \; \mathlarger{V}_{(2)}\big[^{n_{0},n}_{n'_{0},n'}\big]
\label{eq:ST_symmetry2_apndx}
\end{equation}

\newpage
\section{Effective Hamiltonian from perturbation theory}\label{app:Heff}
To obtain an effective Hamiltonian in the subspace spanned by relevant states we use a standard perturbative procedure~\footnote{ See e.g.~ C.~Cohen-Tannoudji, J.~Dupont-Roc and G.~Grynberg; \textit{Atom-photon interactions: basic processes
and applications}; John Wiley \& Sons (1998), Chapter 1 Complement B.}.
We write the Hamiltonian in \ck{Eq.~(4)} as  $\hat{H}_\mathrm{tot}=\hat{H}_{0}+\hat{V}$ where $\hat{V}$ is the second line of \ck{Eq.~(4)} and $V_{(1)}$ and $V_{(2)}$ are given by Eq.(6).
We divide the eigenstates of $\hat{H}_{0}$ into a subspace $\mathcal{P}$ that contains our states of interest and its complement as $\mathcal{Q}$. 
We denote the eigenstates of $H_0$ that are in subspace $\mathcal{P}$ by $\ket{i,\mathcal{P}}$ and the respective eigenenergies by $E_{i}^{(\mathcal{P})}$.
The space $\mathcal{P}$ contains all eigenstates  of $H_0$ such that the for any two states $\{i,j\}$ in subspace $\mathcal{P}$ and a state $k$ in subspace $\mathcal{Q}$, the energy separation obeys $|E_{i}^{\mathcal{(P)}}-E_{j}^{\mathcal{(P)}}| \ll |E_{i}^{\mathcal{(P)}}-E_{k}^{\mathcal{(Q)}}|$.
The matrix elements of the effective Hamiltonian, up to second order in perturbation theory, are then given by
\begin{equation}
\begin{split}
\braket{i,\mathcal{P}|H_{\mathrm{eff}}^{\mathcal{(P)}}|j,\mathcal{P}}
=&\phantom{+}
E_{i}^{\mathcal{(P)}}\delta_{ij}
\\
&+\braket{i,\mathcal{P}|\hat{V}|j,\mathcal{P}}
\\
&+\frac{1}{2}\sum\limits_{k} \braket{i,\mathcal{P}|\hat{V}|k,\mathcal{Q}}\braket{k,\mathcal{Q}|\hat{V}|j,\mathcal{P}}\Big(\frac{1}{E^{\mathcal{(P)}}_{i}-E^{\mathcal{(Q)}}_{k}}+\frac{1}{E^{\mathcal{(P)}}_{j}-E^{\mathcal{(Q)}}_{k}}\Big),
\end{split}
\label{eq:eff_hamil1}
\end{equation}
where $\delta_{ij}$ is the Kronecker delta.

\subsection{Frequency of middle oscillator much larger than those of oscillators 1 and 2. }

Here the relevant state space is spanned by the two states $\ket{1,0,0}$ and $\ket{0,0,1}$, such that  $\mathcal{P}=\Big\{\ket{1,0,0},\ket{0,0,1}\Big\}$ and $\mathcal{Q}=\Big\{\ket{j,k,\ell} \ \mathrm{with}\ \ket{j,k,\ell}\neq\{\ket{1,0,0},\ket{0,0,1}\}\Big\}$.
 In this basis the effective Hamiltonian is
$H_{\mathrm{eff}}^{\mathrm{Dimer}}=\begin{pmatrix}
\epsilon & \kappa\\
\kappa & \epsilon
\end{pmatrix} $
with 
\begin{equation}
\begin{split}    
\kappa
=&
\sum\limits_{j=1}(-1)^{j}
\Bigg(\frac{\mathlarger{V}_{(1)}\big[^{0,0}_{j,1}\big]}{\hbar\omega}\Bigg)^2
\Big(\frac{\omega}{\omega_0}\Big)
\frac{1}{j}
\Bigg[\frac{2 \, }{ 1-(\frac{\omega}{j\omega_0})^2
}\Bigg]\hbar \omega
\label{eq:2state_effective_hamiltonian_app}
\end{split}
\end{equation}
where the symmetries of $\mathlarger{V}_{(1)}\big[^{n'_0,n'_i}_{n_0,n_i}\big]$ have been used, and $\epsilon$ is a global energy shift, that does not influence the dynamics. 
For the perturbation treatment to be valid $\omega_0$ has to be much larger than $\omega$.
Then one can further simplify to arrive at \ck{Eq.~(8)} of the main text.

\newpage
\subsection{All oscillators have roughly the same frequency}

Now the state $\ket{0,1,0}$ is quasi degenerate with the states  $\ket{1,0,0}$ and $\ket{0,0,1}$. 
Therefore we include it in the subspace $\mathcal{P}$. Hence the subspaces now become $\mathcal{P}=\Big\{\ket{1,0,0},\ket{0,1,0},\ket{0,0,1}\Big\}$ and $\mathcal{Q}=\Big\{\ket{j,k,\ell} \ \mathrm{with}\ \ket{j,k,\ell}\neq\{\ket{1,0,0},\ket{0,1,0}, \ket{(0,0,1}\}\Big\}$.
The effective Hamiltonian can, in the basis $\ket{1,0,0}$, $\ket{0,1,0}$, $\ket{0,0,1}$,  be written as
\begin{equation}
\label{eq:3state_effective}
H_{\mathrm{eff}}^{\mathrm{Trimer}}=
\begin{blockarray}{cccc}
\ket{100} & \ket{010} & \ket{001}& \\[2ex]
\begin{block}{(ccc)c}
0&\alpha&\beta&\ket{100}\\[1ex]
\alpha & \Delta & \alpha &\ket{010}\\[1ex]
\beta & \alpha & 0 &\ket{001}
\\
\end{block}
\end{blockarray}
\end{equation}
with detuning

\begin{equation}
    \begin{split}
    \label{eq:Delta}
\Delta
=&\phantom{+}
\hbar(\omega_{0}-\omega)+2\mathlarger{V}_{(1)}\big[^{1,0}_{1,0}\big]-\mathlarger{V}_{(1)}\big[^{0,0}_{0,0}\big]-\mathlarger{V}_{(1)}\big[^{0,1}_{0,1}\big] + \frac{( \mathlarger{V}_{(1)}\big[^{0,0}_{1,0}\big])^{2}}{\hbar\omega_{0}} \\
&+2\sum\limits_{j,k \geq 1}\Big(\frac{( \mathlarger{V}_{(1)}\big[^{1,0}_{j,k}\big])^{2}}{\hbar\omega_{0}(1-j)-k\hbar\omega }\Big) - \sum\limits_{j\geq 2}\Big(\frac{2(-1)^{j}(\mathlarger{V}_{(1)}\big[^{1,0}_{j,0}\big])^{2}}{\hbar\omega_{0}(1-j) }\Big)\\
& - \sum\limits_{j,k \geq 1}\Big(\frac{( \mathlarger{V}_{(1)}\big[^{0,1}_{j,k}\big])^{2}}{-j\hbar\omega_{0}+(1-k)\hbar\omega }\Big) -\sum\limits_{j,\ell \geq 1}\Big(\frac{( \mathlarger{V}_{(1)}\big[^{0,0}_{j,\ell}\big])^{2}}{-j\hbar\omega_{0}-\ell\hbar\omega }\Big) - \sum\limits_{j \geq 1}\Big(\frac{2(-1)^{j}\mathlarger{V}_{(1)}\big[^{0,1}_{j,1}\big] \; \mathlarger{V}_{(1)}\big[^{0,0}_{j,0}\big]}{-j\hbar\omega_{0} }\Big),
\end{split}
\end{equation}
nearest neighbor coupling 
\begin{equation}
\begin{split}
\label{alpha}
\alpha
=
&\mathlarger{V}_{(1)}\big[^{1,0}_{0,1}\big]\\
&+\sum\limits_{j,k \geq 1} \frac{\mathlarger{V}_{(1)}\big[^{1,0}_{j,k}\big] \; \mathlarger{V}_{(1)}\big[^{0,1}_{j,k}\big]}{2 } \; \; \; \Big(\frac{1}{\hbar\omega_{0}(1-j)-k\hbar\omega}+\frac{1}{-j\hbar\omega_{0}+\hbar\omega(1-k)}\Big)\\
& -\sum_{j\geq 2} (-1)^{j}\frac{\mathlarger{V}_{(1)}\big[^{0,1}_{j,0}\big] \; \mathlarger{V}_{(1)}\big[^{1,0}_{j,0}\big]}{2} \; \; \; \Big(\frac{1}{\hbar\omega_{0}(1-j)}+\frac{1}{-j\hbar\omega_{0}+\hbar\omega}\Big)\\
&+
\sum_{j\ge 1} (-1)^{j}\frac{\mathlarger{V}_{(1)}\big[^{1,0}_{j,1}\big] \; \mathlarger{V}_{(1)}\big[^{0,0}_{j,0}\big]}{2} \; \; \; \Big(\frac{1}{\hbar\omega_{0}(1-j)-\hbar\omega}+\frac{1}{-j\hbar\omega_{0}}\Big),
\end{split}
\end{equation}
and direct coupling between the states $\ket{1,0,0}$ and $\ket{0,0,1}$: 
\begin{equation}
\begin{split}
\beta=
&\sum\limits_{j=0}(-1)^{(j+1)}\frac{\big(\mathlarger{V}_{(1)}\big[^{0,0}_{j,1}\big]\big)^{2} }{-j\hbar\omega_{0}-\hbar\omega} - \sum\limits_{j=0,j\neq 1}(-1)^{(j)}\frac{\big(\mathlarger{V}_{(1)}\big[^{0,1}_{j,0}\big]\big)^{2} }{-j\hbar\omega_{0}+\hbar\omega}  \\
=& -\Big(\frac{1}{\hbar\omega_{0}}\Big)\frac{\big(\mathlarger{V}_{(1)}\big[^{0,0}_{1,1}\big]\big)^{2} }{1+\frac{\omega}{\omega_{0}}}  - \sum\limits_{j=2}(-1)^{j}\big(\mathlarger{V}_{(1)}\big[^{0,0}_{j,1}\big]\big)^{2} \Big[ \frac{-2j\hbar\omega_{0}}{j^{2}\hbar^{2}\omega_{0}^{2}-\hbar^{2}\omega^{2}}  \Big] \\
=&  \Bigg( -\Big(\frac{\mathlarger{V}_{(1)}\big[^{0,0}_{1,1}\big] }{\hbar\omega}\Big)^{2} \Big(\frac{\omega}{\omega_{0}}\Big)\Big[ \frac{1}{1+\big(\frac{\omega}{\omega_{0}}\big)}  \Big] + \sum\limits_{j=2}(-1)^{j}\Big(\frac{\mathlarger{V}_{(1)}\big[^{0,0}_{j,1}\big] }{\hbar\omega}\Big)^{2} \Big(\frac{\omega}{\omega_{0}}\Big) \frac{1}{j}\Big[ \frac{2}{1-\big(\frac{\omega}{j\omega_{0}}\big)^{2}}  \Big] \Bigg) \hbar\omega.
\end{split}
\label{eq:beta_jsum}
\end{equation}

\subsection{Derivation of $\alpha\approx-\Delta$ for $\omega_{0}\approx \omega$ }

For the case when $\omega_{0}=\omega$, the $\mathlarger{V}_{(1)}\big[^{n_{0},n_{i}}_{n'_{0},n'_{i}}\big]$ can be simplified. 
We begin by writing the product of Harmonic oscillator wavefunctions in the integrand of \ck{Eq.~(6)}  as
\begin{equation}
\mathcal{I}=\psi_{n_{0}}\Big(\frac{R+\delta-D}{2}\Big)\;\psi_{n_{i}}\Big(\frac{R-\delta+D}{2}\Big)\;\psi_{n'_{0}}\Big(\frac{R+\delta-D}{2}\Big)\;\psi_{n'_{i}}\Big(\frac{R-\delta+D}{2}\Big)\\
\end{equation}
where we have used transformed coordinates $\delta=(D-x_{1}+x_{0})$ and $R=x_{1}+x_{0}$. 
The back transformation is given as $x_{1}=\frac{R-\delta+D}{2}$ and $x_{0}=\frac{R+\delta-D}{2}$. 

Expressing the oscillator functions in terms of Hermite polynomials $H_n$ we obtain
\begin{equation*}
\begin{split}
\mathcal{I}
=&\sqrt{\frac{1}{2^{\big(n_{0}+n_{i}+n'_{0}+n'_{i}\big)}}} \Big(\frac{1}{\pi L^{2}}\Big) \times e^{-\frac{R^{2}}{2L^{2}}} \times e^{-\frac{(\delta-D)^{2}}{2L^{2}}}\\
&\times H_{n_{0}}\Big(\frac{R+(\delta-D)}{2L}\Big)H_{n'_{0}}\Big(\frac{R+(\delta-D)}{2L}\Big) \; H_{n_{i}}\Big(\frac{R-(\delta-D)}{2L}\Big)H_{n'_{i}}\Big(\frac{R-(\delta-D)}{2L}\Big),\\
&
\end{split}
\end{equation*}
where $L=\sqrt{\frac{\hbar}{m\omega}}$. 
 In leading order of the interaction Eq.~(\ref{eq:Delta}) and Eq.~(\ref{alpha})  are
\begin{equation}
    \Delta\approx\; \hbar(\omega_{0}-\omega) +2\mathlarger{V}_{(1)}\big[^{1,0}_{1,0}\big]-\mathlarger{V}_{(1)}\big[^{0,0}_{0,0}\big]-\mathlarger{V}_{(1)}\big[^{0,1}_{0,1}\big]
\quad\mathrm{and}
\quad
\alpha=\mathlarger{V}_{(1)}\big[^{1,0}_{0,1}\big]
\end{equation}
where
\begin{equation}
\begin{split}
&\mathlarger{V}_{(1)}\big[^{0,0}_{0,0}\big]=\frac{U_{e}}{\pi L^{2}}  \iint  d\delta \; dR   \; \; \; e^{-\frac{R^{2}}{2L^{2}}}  e^{-\frac{(\delta-D)^{2}}{2L^{2}}} 
\; \; \; \Big|\phi_{\nu}^{(\mathrm{s})}\big(|\delta| \big)\Big|^{2}\\
&\mathlarger{V}_{(1)}\big[^{0,1}_{0,1}\big]=\frac{U_{e}}{2\pi L^{4}}  \iint  d\delta \; dR   \; \; \Big(R-(\delta-D)\Big)^{2}  e^{-\frac{R^{2}}{2L^{2}}}  e^{-\frac{(\delta-D)^{2}}{2L^{2}}} 
\; \; \; \Big|\phi_{\nu}^{(\mathrm{s})}\big(|\delta| \big)\Big|^{2}\\
&\mathlarger{V}_{(1)}\big[^{1,0}_{1,0}\big]=\frac{U_{e}}{2\pi L^{4}}  \iint  d\delta \; dR   \; \; \Big(R+(\delta-D)\Big)^{2}  e^{-\frac{R^{2}}{2L^{2}}}  e^{-\frac{(\delta-D)^{2}}{2L^{2}}} 
\; \; \; \Big|\phi_{\nu}^{(\mathrm{s})}\big(|\delta| \big)\Big|^{2}
\\
&\mathlarger{V}_{(1)}\big[^{1,0}_{0,1}\big]=\frac{U_{e}}{2\pi L^{4}}  \iint  d\delta \; dR   \; \; \Big(R-(\delta-D)\Big) \Big(R+(\delta-D)\Big)  e^{-\frac{R^{2}}{2L^{2}}}  e^{-\frac{(\delta-D)^{2}}{2L^{2}}} \; \; \; \Big|\phi_{\nu}^{(\mathrm{s})}\big(|\delta| \big)\Big|^{2}.
\end{split}
\label{eq:Tijkl_1}
\end{equation}
Replacing $R\rightarrow -R$, it is easy to see that $\mathlarger{V}_{(1)}\big[^{0,1}_{0,1}\big]=\mathlarger{V}_{(1)}\big[^{1,0}_{1,0}\big]$. Therefore we shall consider only $\mathlarger{V}_{(1)}\big[^{1,0}_{1,0}\big]$ below.

After performing the integration  over the variable $R$, we obtain
\begin{equation}
\begin{split}
&\mathlarger{V}_{(1)}\big[^{0,0}_{0,0}\big]=\frac{U_{e}}{L}\sqrt{\frac{2}{\pi}}  \int  d\delta \;\;\; e^{-\frac{(\delta-D)^{2}}{2L^{2}}} \Big|\phi_{\nu}^{(\mathrm{s})}\big(|\delta| \big)\Big|^{2},\\
& \\
&\mathlarger{V}_{(1)}\big[^{1,0}_{1,0}\big]=\frac{U_{e}}{2L}\sqrt{\frac{2}{\pi}} \int  d\delta \;\;\; e^{-\frac{(\delta-D)^{2}}{2L^{2}}} \Big|\phi_{\nu}^{(\mathrm{s})}\big(|\delta| \big)\Big|^{2} \;\;\; + \;\;\;\frac{U_{e}}{2L^{3}}\sqrt{\frac{2}{\pi}}\int  d\delta \; \; (\delta-D)^{2} \times e^{-\frac{(\delta-D)^{2}}{2L^{2}}} \Big|\phi_{\nu}^{(\mathrm{s})}\big(|\delta| \big)\Big|^{2},\\
& \\
&\mathlarger{V}_{(1)}\big[^{1,0}_{0,1}\big]=\frac{U_{e}}{2L}\sqrt{\frac{2}{\pi}} \int  d\delta \;\;\; e^{-\frac{(\delta-D)^{2}}{2L^{2}}} \Big|\phi_{\nu}^{(\mathrm{s})}\big(|\delta| \big)\Big|^{2}\;\;\; - \;\;\;\frac{U_{e}}{2L^{3}}\sqrt{\frac{2}{\pi}}\int  d\delta \; \; (\delta-D)^{2} e^{-\frac{(\delta-D)^{2}}{2L^{2}}} \Big|\phi_{\nu}^{(\mathrm{s})}\big(|\delta| \big)\Big|^{2}.
\end{split}
\label{eq:Tijkl_3}
\end{equation}
With these expressions for the matrix elements, we write $\Delta$ to leading order as
\begin{equation}
\begin{split}
\Delta\approx&\ \hbar(\omega_{0}-\omega)+\Big[\frac{U_{e}}{2L}\sqrt{\frac{2}{\pi}} \int  d\delta \;\;\; e^{-\frac{(\delta-D)^{2}}{2L^{2}}} 
	\Big|\phi_{\nu}^{(\mathrm{s})}\big(|\delta| \big)\Big|^{2} \;\;\; \\
&\hspace{0cm}+ \;\;\; \frac{U_{e}}{2L^{3}}\sqrt{\frac{2}{\pi}}\int  d\delta \; \; (\delta-D)^{2} \times e^{-\frac{(\delta-D)^{2}}{2L^{2}}} \Big|\phi_{\nu}^{(\mathrm{s})}\big(|\delta| \big)\Big|^{2}\Big]- \Big[\frac{U_{e}}{L}\sqrt{\frac{2}{\pi}} \int  d\delta \;\;\; e^{-\frac{(\delta-D)^{2}}{2L^{2}}} 
	\Big|\phi_{\nu}^{(\mathrm{s})}\big(|\delta| \big)\Big|^{2}\Big]\\
&\\
=&\ \hbar(\omega_{0}-\omega)\\
&\ -\Big[\frac{U_{e}}{2L}\sqrt{\frac{2}{\pi}} \int  d\delta \;\;\; e^{-\frac{(\delta-D)^{2}}{2L^{2}}} \Big|\phi_{\nu}^{(\mathrm{s})}\big(|\delta| \big)\Big|^{2}\;\;\; - \;\;\; \frac{U_{e}}{2L^{3}}\sqrt{\frac{2}{\pi}}\int  d\delta \; \; (\delta-D)^{2} \times e^{-\frac{(\delta-D)^{2}}{2L^{2}}}  \Big|\phi_{\nu}^{(\mathrm{s})}\big(|\delta| \big)\Big|^{2}\Big]\\
&\\
= &\ \hbar(\omega_{0}-\omega)-\mathlarger{V}_{(1)}\big[^{1,0}_{0,1}\big] 
\\
\approx&\ \hbar(\omega_{0}
-\omega)-\alpha.
\end{split}
\end{equation}

\vspace{2cm}

\clearpage
\newpage

\clearpage
\newpage
\section{Comparison between perturbation theory and numerical results}

\begin{figure}[b]
    \centering
    \includegraphics[width=\columnwidth]{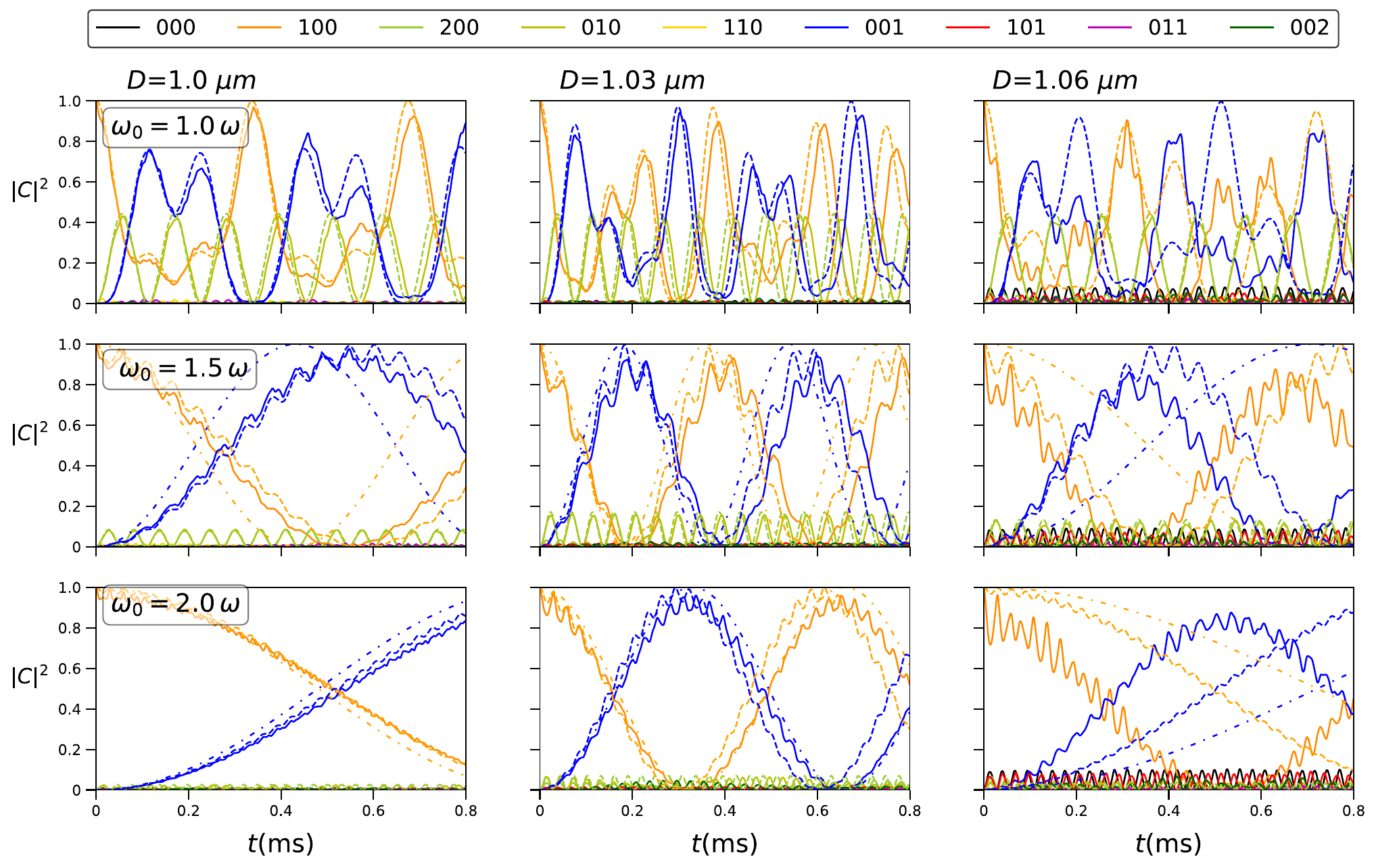}
    \caption{Comparison of the  population dynamics of the effective Hamiltonians with the  full Hamiltonian. 
    The parameters of $D$ and $\omega_0$ are the same as Fig.~2 of the main text. 
    Solid line: full solution,
     dashed line: Eq.~(\ref{eq:3state_effective}), dashdot line: Eq.~(\ref{eq:2state_effective_hamiltonian_app}).
}
    \label{appendix_fig:fig2}
\end{figure}

\begin{figure}[p]
    \centering
    \includegraphics[width=\columnwidth]{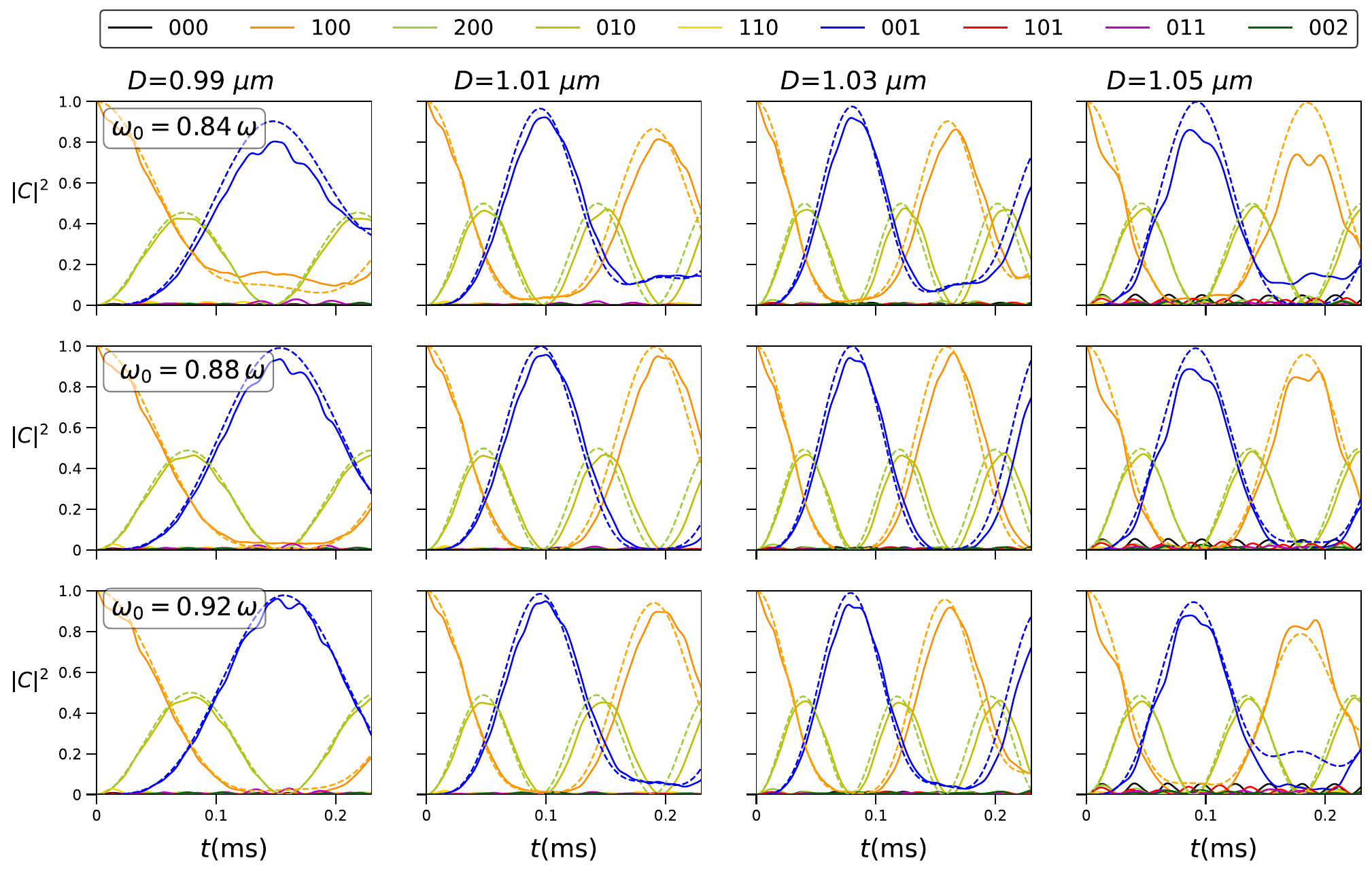}
    \caption{Comparison of the  population dynamics of the effective Hamiltonians Eq.~(\ref{eq:3state_effective}) with the  full Hamiltonian. 
    The parameters of $D$ and $\omega_0$ are the same as \ck{Fig.~3} of the main text. 
    Solid line: full solution,
     dashed line: Eq.~(\ref{eq:3state_effective}).}
    \label{appendix_fig:fig3}
\end{figure}

\begin{figure}[p]
    \centering
    \includegraphics[width=\columnwidth]{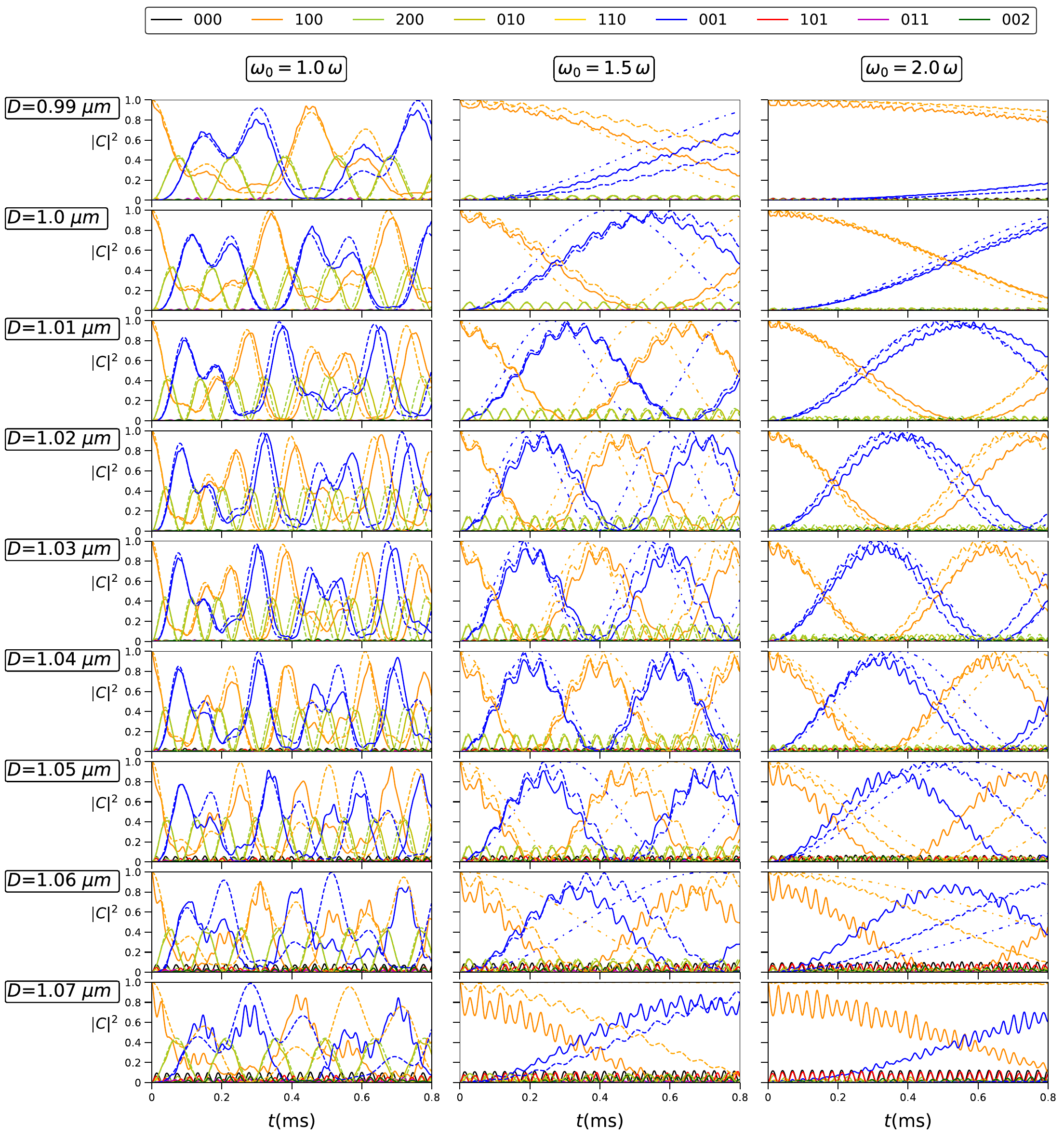}
    \caption{Comparison of the  population dynamics of the effective Hamiltonians with the  full Hamiltonian. 
    The parameters  $\omega_0$ are the same as Fig.~2 of the main text. 
    Solid line: full solution,
     dashed line: Eq.~(\ref{eq:3state_effective}), dashdot line: Eq.~(\ref{eq:2state_effective_hamiltonian_app}).
     Note that the arrangement of rows and columns is different from that of the main text.
}
    \label{appendix_fig:fig2long}
\end{figure}

\begin{figure}[p]
    \centering
    \includegraphics[width=\columnwidth]{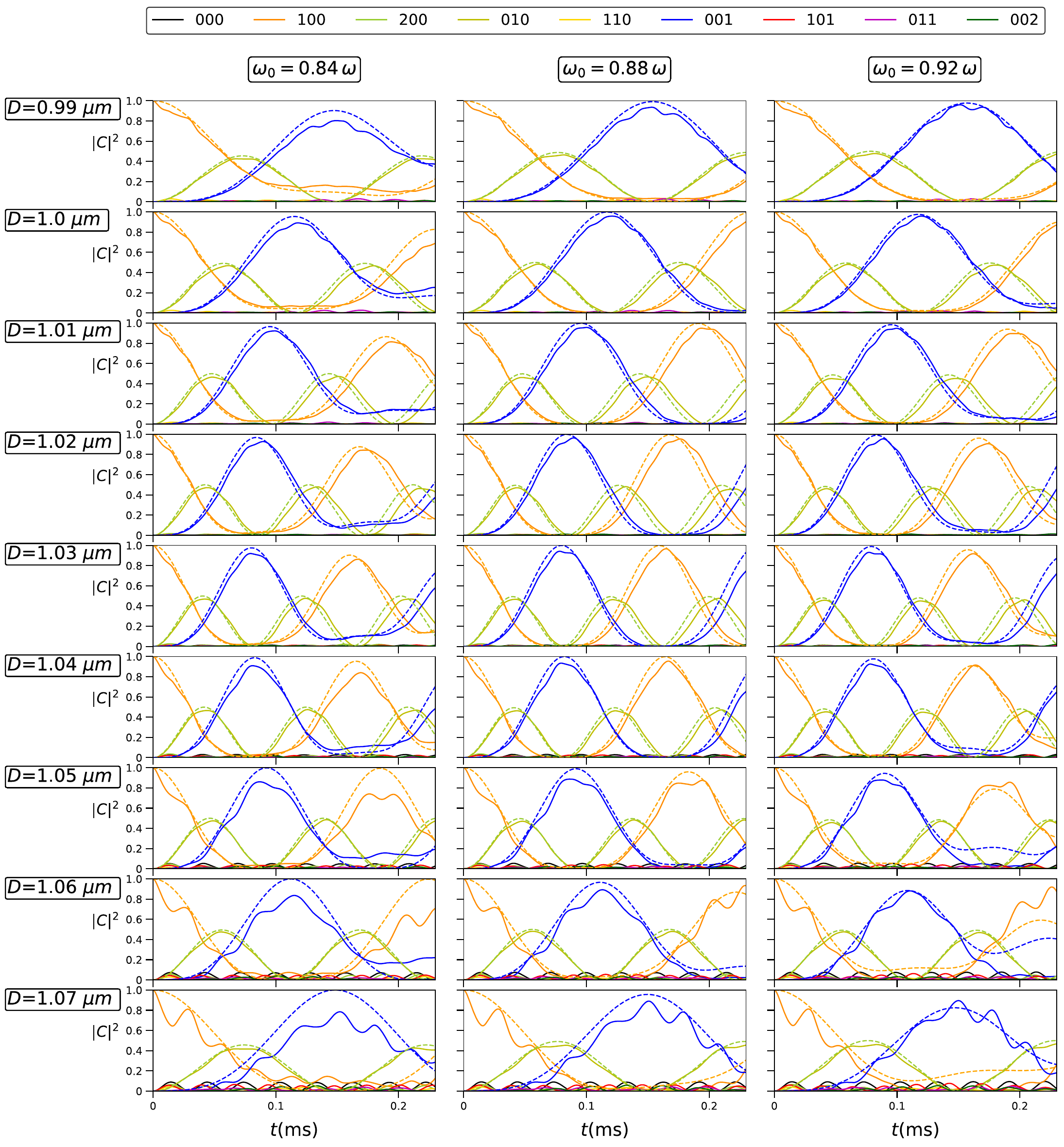}
    \caption{Comparison of the  population dynamics of the effective Hamiltonian Eq.~(\ref{eq:3state_effective}), dashed line,  with the  full Hamiltonian (solid line). 
    The parameters of  $\omega_0$ are the same as Fig.~3 of the main text. 
    Note that the arrangement of rows and columns is different from that of the main text.
}
    \label{appendix_fig:fig3long}
\end{figure}

In the following figures a comparison between the population dynamics for full Hamiltonian with the population dynamics obtained using the effective Hamiltonians is shown. 
In Figs.~\ref{appendix_fig:fig2} and \ref{appendix_fig:fig2long} we use the same frequencies as 
in Fig.~2 of the main text (i.e., $ \omega_0= \omega$, $\omega_0=1.5\,\omega$ and $\omega_0=2\,\omega$).
In Figs.~\ref{appendix_fig:fig3} and \ref{appendix_fig:fig3long} we use the same frequencies as 
in Fig.~4 of the main text (i.e., $ \omega_0= 0.84\,\omega$, $\omega_0=0.88\,\omega$ and $\omega_0=0.92\,\omega$).

\newpage

